# O VI Absorbers Tracing Hot Gas Associated with a Pair of Galaxies at z = 0.167[1]

B. D. Savage[2], A. Narayanan[2], B. P. Wakker[2], J. T. Stocke[3], B. A. Keeney[3],

J. M. Shull[3], K. R. Sembach[4], Y. Yao[3], C., & J. Green[3]

ABSTRACT

High signal-to-noise (S/N) observations of the QSO PKS 0405-123 ($z_{em}$ = 0.572) with the Cosmic Origins Spectrograph from 1134 to 1796 Å with a resolution of ~17 km s$^{-1}$ are used to study the multi-phase partial Lyman limit system (LLS) at z = 0.16716 which has previously been studied using relatively low S/N spectra from STIS and FUSE. The LLS and an associated H I-free broad O VI absorber likely originate in the circumgalactic gas associated with a pair of galaxies at z = 0.1688 and 0.1670 with impact parameters of 116 $h_{70}^{-1}$ and 99 $h_{70}^{-1}$. The broad and symmetric O VI absorption is detected in the z = 0.16716 restframe with v = -278±3 km s$^{-1}$, log N(O VI) = 13.90±0.03 and b = 52±2 km s$^{-1}$. This absorber is not detected in H I or other species with the possible exception of N V. The broad, symmetric O VI profile and absence of corresponding H I absorption indicates that the circumgalactic gas in which the collisionally ionized O VI arises is hot (log T ~ 5.8-6.2). The absorber may represent a rare but important new class of low z IGM absorbers. The LLS has strong asymmetrical O VI absorption with log N(O VI) = 14.72±0.02 spanning a velocity range from -200 to +100 km s$^{-1}$. The high and low ions in the LLS have properties resembling those found

---

for Galactic highly ionized HVCs where the O VI is likely produced in the conductive and turbulent interfaces between cool and hot gas.

*Short title:* O VI Absorbers Tracing Hot Gas

1. INTRODUCTION

Ultraviolet absorption line studies of the low redshift IGM with the Hubble Space Telescope have revealed that ~50% of the baryons reside in gas with $10^4 < T < 10^6$ K. The cooler baryons with $T \sim (2-4) \times 10^4$ K are traced by the Lyman $\alpha$ forest (Penton et al. 2004; Lehner et al. 2007; Danforth & Shull 2008) and some of the O VI lines (Tripp et al. 2008; Thom & Chen 2008). The warmer baryons with $T \sim (1-10) \times 10^5$ K have been traced by broad Ly$\alpha$ (Richter et al. 2006; Lehner et al. 2007; Danforth et al. 2010a), O VI absorption lines (Danforth & Shull 2008; Tripp et al. 2008) and Ne VIII absorption (Savage et al. 2005; Narayanan et al. 2009). The number of baryons detected in the IGM in the UV exceeds the ~10% revealed by luminous matter (stars and galaxies) by about a factor of 6. The remaining ~40% of the baryons likely reside in hotter gas with $T > 10^6$ K. The hot baryons can be studied at X-ray wavelengths through searches for O VII and O VIII and other X-ray lines (for a review see Bregman 2007). At UV wavelengths hot baryons can be revealed by redshifted EUV absorption lines of Ne VIII, Mg X and Si XII or by broad but weak absorption by O VI and H I.

The baryons in the IGM play an important role in galaxy formation and evolution. At low z it is possible to probe the direct association of the intergalactic baryons with galaxies through observations of the redshifts of galaxies in the general directions of bright QSOs observed in the ultraviolet. Those studies have shown that ~50% of the low redshift H I absorption lines are tracing gas in the environments of galaxies out to distances of ~500 kpc (Morris et al. 1993; Tripp et al.1998; Penton et al. 2002; Stocke et al. 2006; Chen & Mulchaey 2009; Wakker & Savage 2009) Hydrodynamical simulations of the assembly of matter in the evolving universe predict the formation of a Cosmic Web of filamentary structures with gas falling into gravitational potential wells that can be shock heated to $T>10^6$ K (Cen & Ostriker 1999; Davé et al. 2001; Fang & Bryan 2001; Oppenheimer & Davé 2008). The galaxies forming from these processes may have surrounding hot halos or be in gaseous environments that connect to hot filament gas. The feedback from galaxies in the form of large scale outflows can heat and enrich the galaxy surroundings, providing additional processes for the creation of highly ionized circumgalactic plasmas (Cen & Fang 2006; Oppenheimer & Davé 2009). Studying the gas/galaxy interface regions may provide important insights about continuing galaxy formation processes, galaxy outflows and infall and the interactions of galaxies with their surroundings.

The Cosmic Origins Spectrograph (COS) has opened a new era in ultraviolet studies of the low redshift IGM. Although COS has a spectral resolution (FWHM) ~2.5 times lower than the E140M mode on the Space Telescope Imaging Spectrograph (STIS), COS is 10-20 times more efficient in collecting photons, making it possible to obtain relatively high signal-to-noise (S/N) spectra of bright QSOs for absorption line studies of the IGM. With high S/N observations, it is possible to exploit the full diagnostic power of UV absorption-line spectroscopy. Very important spectroscopic features are weak and broad and difficult to detect in low S/N spectra. In this paper we report the discovery of a weak and broad O VI absorption line displaced -278 km s$^{-1}$ from a Lyman limit absorption system at z = 0.16716 in the spectrum of PKS 0405-123, a QSO with $z_{em}$ = 0.572. The broad O VI absorption has no associated H I and may represent the direct detection of hot, T~ (1-3)x10$^6$ K gas associated with a foreground galaxy. The Lyman limit absorber has multi-phase gas properties similar to the highly ionized high velocity clouds (HVCs) surrounding the Milky Way. The O VI associated with the Lyman limit absorber is probably produced in gas in the interface regions between the cool gas of the HVC and surrounding hot gas associated with a foreground galaxy or the group in which the galaxies reside.

This paper is organized as follows. In §2 we discuss the COS observations and the data processing methods. In §3 we describe the properties of the two galaxies with impact parameters <120 kpc associated with the absorbers near z = 0.16716. §4 contains an analysis of the properties of the strong Lyman limit absorber at z = 0.16716 including discussions of the origin of the ionization in the different gas phases detected. §5 discusses the properties and ionization of the broad H I-free O VI absorber displaced -278 km s$^{-1}$ from the Lyman limit absorber. The possible physical origins of the Lyman limit absorber and the broad H I-free O VI absorber are discussed in §6. The overall results of this investigation are summarized in §7.

Distances in this paper are physical distances assuming a ΛCDM cosmology with $\Omega_M$ = 0.3, Λ = 0.7 and $H_0$ = 70 km s$^{-1}$ Mpc$^{-1}$ with $h_{70}$ = $H_0$/ 70 km s$^{-1}$ Mpc$^{-1}$.

## 2. COS OBSERVATIONS OF PKS 0405-123

We have combined the COS G130M and G160M observations of PKS 0405-123 spanning the wavelength range from 1134 to 1796 Å listed in Table 1. The August 2009 G130M observations were part of the HST Early Release Program. The December 2009 G130M and G160M observations are from the COS science team program. Information about COS can be found in Froning & Green (2009), Green et al. (2010), and the COS HST Instrument Handbook (Dixon et al. 2010). The inflight performance of COS is discussed in Osterman et al. (2010) and in numerous instrument science reports on the STScI COS website at http://www.stsci.edu/hst/cos/documents/isrs.

Spectral integrations were obtained with different grating set up wavelengths in order to obtain spectra with different detector/wavelength alignments to reduce the effects of detector fixed pattern noise and provide a way of covering the wavelength gaps between the two detector micro-channel plate segments (A and B). Table 1 lists the various integrations and their HST MAST identification code, dates of observation, and integration times. We did not include in our analysis integrations obtained 3 August 2009 before the spectrograph was at optimum focus.

The micro-channel plate delay line detector was operated in the time-tag mode with the QSO centered in the 2.5" diameter primary science aperture. The internal wavelength calibration lamps were flashed several times during each science integration. The time-tag data were processed with CalCOS with the reference files available 1 January 2010.

The detector and scattered light background level in COS spectra is extremely small. Flat-fielding, alignment and co-addition of the individual processed spectra utilized software developed by the COS team for the processing of FUV observations.[1] The effects of major detector defects were removed when producing the combined spectrum by giving the affected wavelength regions in each individual spectrum low weight during the addition process . The reduced intensity in the grid wire shadows was also corrected for in the individual integrations and these affected wavelength regions were given lower weight in the co-addition process. The extraction process is explained in more detail in Danforth et al. (2010b).

The proper alignment of the individual spectra was achieved through a cross-correlation technique. The different individual spectra (in flux units) were weighted by integration time when combined.

An example of the combined COS G130M spectrum from $\lambda$1195 to 1240 Å is shown in the lower panel of Figure 1. The total integration time is 14.5 ksec. For comparison, the STIS spectrum with the E140M echelle grating at a resolution of ~7 km s$^{-1}$ obtained with a 27 ksec integration is shown in the upper panel. With a photon detection efficiency of ~15 times higher than STIS, COS is able to produce spectra with much higher S/N but at several times lower resolution. IGM absorption lines are identified above the COS spectrum. Two O VI $\lambda\lambda$1032, 1038 systems and their associated H I Ly $\beta$ $\lambda$1025 at z = 0.16716 and 0.18295 are marked. They are clearly seen in the COS spectrum but detected at much lower significance in the STIS spectrum. The system at z = 0.16716 is the major topic of this paper.

---

[1] See http://casa .colorado.edu/danforth/costools.html for the COS team co-addition and flat-fielding software and for additional discussions.

The S/N in the resulting combined spectrum associated with photon counting ranges from 35 to 50 for 1150 < $\lambda$ < 1750 Å. However, since the weaker fixed pattern noise features were not flat fielded but simply averaged, there are residual features at the 2-3% level throughout the spectrum. The residual features are stronger near the ends of the detector segments. We note that the glass fiber bundles that compose the micro-channel plates in the detector are packaged in a hexagon pattern which introduces a repeating fixed pattern noise structure in individual integrations with a peak-to-peak amplitude of ~10%. Averaging over these structures with multiple integrations at different set up wavelengths reduces their amplitude to ~2-3 %.

A preliminary characterization of the COS line spread function (LSF) is found in Ghavamian et al. (2009). The LSF has a narrow core and broad wings, well described by narrow and broad Gaussian functions. At 1200, 1300, and 1400 Å the LSF has a full width at half maximum (FWHM) of 0.0684, 0.0669, and 0.0648 Å implying velocity widths of 17.1, 15.4, and 13.9 km s$^{-1}$, respectively. However, the broad wings on the LSF have FWHM ~50 km s$^{-1}$ and contain ~20 to 30% of the LSF area. The broad wing contribution is largest at the shortest wavelengths. Ghavamian et al. (2009) provide a number of examples of the impact of the broad wings on complex absorption line spectra containing strong and weaker lines by comparing COS and STIS 2.6 km s$^{-1}$ (FWHM) observations of interstellar absorption toward SK 155 in the Small Magellanic Cloud. When a large number of independent integrations are combined, there may be an increase of the width of the core of the spread function due to slight amounts of wavelength misalignment from one integration to the next.

Wavelengths, velocities and redshifts reported in this paper are heliocentric. The COS wavelength calibration is obtained through Pt lamp exposures obtained as part of each integration. Radio observations of H I 21 cm emission in the direction of PKS 0405-123 with the 140' Greenbank telescope from Lockman & Savage (1995) reveal multi-component H I emission extending from $v_{LSR}$ = -53 to 39 km s$^{-1}$ with an average LSR velocity of 3.0 km s$^{-1}$. Toward PKS 0405-123 $v_{HELIO} - v_{LSR}$ = +16.9 km s$^{-1}$. We therefore would expect the weak to moderately strong ISM absorption lines of neutral gas tracers observed by COS to be centered at $v_{HELIO}$ = 20 km s$^{-1}$. Weak low ionization ISM lines observed in the G130M and G160M integrations were used to check the reliability of the COS velocity calibration. We applied corrections of 6 km s$^{-1}$ to the G130M observations and 22 km s$^{-1}$ to the G160M observations in order to make the ISM lines have average heliocentric velocities of 20 km s$^{-1}$. Residual errors in the wavelength calibration correspond to velocity errors of ~5 km s$^{-1}$.

## 3. GALAXIES ASSOCIATED WITH THE ABSORBERS NEAR z = 0.16716

The galaxies associated with the strong multi-phase partial Lyman Limit absorber at z = 0.16716 have been identified by Spinrad et al. (1993), Prochaska et al. (2006), and Chen & Mulchaey (2009). There are two galaxies near the absorbers having apparent magnitudes R = 17.4 and 21.0 within $\Delta v$ < 100 km s$^{-1}$ and impact parameter $\rho$ <120$h_{70}^{-1}$ physical kpc of the line of sight to the QSO. No additional galaxies are found near the QSO sightline, although the completeness of the galaxy surveying in this region is variable. For example, the latest study (Chen & Mulchaey 2009) reports a >80% completeness to R=22 within 330 $h_{70}^{-1}$ and >50% completeness within 3.0 $h_{70}^{-1}$ Mpc of the absorber to R = 20 which corresponds to $L_R \sim 0.4L^*$ at z = 0.167. Figure 2 displays a high resolution HST WFPC2 F702W image adapted from Chen & Mulchaey (2009) showing the two known associated galaxies and their positions with respect to PKS 0405-123. Even though the above quoted completeness levels are adequate for statistical cross-correlation studies, missing only ~20% of galaxies associated with the z=0.167 absorber could lead to some misinterpretation of the physical nature of this one absorber complex. For example, there are several galaxies to the east and southeast of the QSO with apparent magnitudes comparable to the z=0.1670 dwarf galaxy to the north, but with no measured redshifts. If some of these are at z=0.167, then the absorbers lie close on the sky to the center of a small group of galaxies.

Enlarged views of the two galaxies are shown in Chen & Mulchaey (2009; see their Fig. 10). The two galaxies have the following properties:

(1) $\rho = 116\ h_{70}^{-1}$ kpc, z = 0.1668±0.0003, R = 17.43, $M_R$ = -21.7, a luminous 4.4$L^*$ galaxy at RA(J2000) = 04h 07m 21.20s and Dec(J2000) = -12º 11' 38.0" identified as object 1753 by Prochaska et al. (2006). As Figure 2 shows, this Sb spiral has compact H II regions along its spiral arms and maybe either face-on or at intermediate inclination; we cannot be more definitive because its image is cut off by the chip edge. The optical spectrum from 3500 to 6500 Å is shown by Prochaska et al. (2006) in their Fig. 8. The spectrum shows strong Ca H+K, G-band and Balmer absorption lines and weak [O II] emission. No [O III] λ5007 emission is evident and Hα is outside the spectral coverage. Prochaska et al. suggest that the Balmer absorption implies a significant episode of star formation ~1 Gyr ago. The galaxy has a systemic velocity of -93±77 km s$^{-1}$ with respect to the absorber reference redshift of 0.16716.

(2) $\rho = 99$ $h_{70}^{-1}$ kpc, z = 0.1670±0.0003, R = 21.00, $M_R$ = -18.9, corresponding to an 0.15$L^*$ dwarf galaxy (ID 90033) with modest-strength emission lines and a disk-like morphology that maybe mildly disturbed by a faint companion (Prochaska et al. 2006). The galaxy has a systemic velocity of -48±77 km$^{-1}$ with respect to the reference absorber redshift.

Note that the reported redshifts for the luminous galaxy are 0.1667 (Spinrad et al. 1993), 0.1668 (Chen & Mulchaey 2009) and 0.1670 (Prochaska et al. 2006). We have adopted the value z = 0.1668 and assign 1$\sigma$ measuring errors to both galaxy redshifts of 0.0003 as suggested by Chen & Mulchaey (2009).

The spiral and dwarf galaxies have a projected separation of 156 $h_{70}^{-1}$. It is possible the two galaxies share a common halo or that the dwarf galaxy moves through the halo of the luminous galaxy producing a tidal tail of gas.

If the absence of other galaxies to $M_R \sim -17$ within $\rho = 1$ $h_{70}^{-1}$ Mpc is confirmed by further redshift survey work to the east and SE, this would imply that the line of sight does not go through a galaxy group, and the absorbers are very likely associated with the gaseous surroundings of one or both of the two foreground galaxies. If higher accuracy galaxy redshifts confirm them both to be at velocities in between the strong H I absorber velocities reported in previous studies (see §4 and Figure 3 below), then a dynamic origin (e.g., outflow or infall) for these absorbers with respect to these galaxies is suggested. The hotter, H I-free absorber reported here for the first time in §5 is at a significantly lower velocity than either of these two galaxies, which makes a simple interpretation of this absorber difficult (see §6.2).

In recent studies, H-W Chen (Chen et al. 2005; Chen & Mulchaey 2009) has stressed that strong Lyα absorbers (log N(H I) ~ 14) correlate almost exclusively with emission-line galaxies, suggesting that the nearby absorbers could arise in the outflows created by intense star formation or in the infall that feeds new star-forming episodes. Even more recently, Menard et al. (2010) has shown that there is an exceptionally strong correlation between the strength of Lyman limit absorbers (log N(H I) ≥ 17; i.e. somewhat thicker than the absorbers studied here) and associated galaxy [O II] emission line strength (and thus current star formation). This extends Chen's results to higher column densities and reinforces the possibility for the absorber association with individual galaxies as infall or outflow. Given these results, the absorbers near $z = 0.16716$ may plausibly trace gas associated either with the luminous spiral galaxy or the dwarf galaxy, both of which have detected emission lines.

## 4. PROPERTIES AND IONIZATION OF THE $z = 0.16716$ MULTI-PHASE ABSORBER

In this section we discuss the properties and ionization of the multi-phase O VI absorber at $z = 0.16716$ in the spectrum of PKS 0405-123. The absorber has been detected by COS and STIS in a wide range of ions including O VI, N V, C IV, C III, C II, N III, N II, Si IV, Si III, Si II, Fe III, O I, and H I. The FUSE + STIS observations cover the wavelength range from 912 to 1750 Å with S/N ~ 5-20, while the COS observations cover 1134 to 1796 Å with S/N ~ 25 to 50. We first describe the essential results from the earlier detailed analysis of the FUSE+STIS observations (§4.1) and then discuss the new insights provided by the higher S/N COS observations (§4.2). We discuss the ionization of the moderately ionized gas in §4.3 and the highly ionized gas in §4.4.

The system has a complex velocity structure, with absorption in the various ions spanning a full velocity range of ~200 km s$^{-1}$, with maximum absorption in O VI close to z = 0.16716. When discussing the velocity structure and displaying absorption line velocity profiles, we adopt the redshift z = 0.16716, where O VI has maximum absorption, as the reference redshift for the system. Continuum-normalized absorption line profiles versus heliocentric velocity for the COS observations are shown in Figure 3. The STIS line profiles are shown in Prochaska et al. (2004, see Figs. 6, 9 and 10). The spectrum of PKS 0405-123 is complex. This produces a number of lines unrelated to the system at z = 0.16716 in the various panels of Figure 3. These unrelated lines are numbered in Figure 3. Their identifications are as follows: (1) C II $\lambda$1036 at z = 0.16716; (2) O VI $\lambda$1038 at z = 0.16716; (3) H I $\lambda$1216 at z = 0.19255; (4) Si IV $\lambda$1394 at z = 0; (5) H I $\lambda$1216 at z = 0.21071; (6) H I $\lambda$1216 at z = 0.25155; (7) O VI $\lambda$1038 at z = 0.3633; (8) C II $\lambda$1335 at z = 0; (9) H I $\lambda$1025 at z = 0.40571; (10) C II$^*$ $\lambda$1336 at z = 0; (11) H I $\lambda$931 at z = 0.40571; (12) H I $\lambda$1216 at z = 0.07677; (13) P II $\lambda$1153 at z = 0. The identifications were made with reference to Williger et al. (2006) and Lehner et al (2007). We do not agree with the Williger et al (2006) identification of line (12) in the Fe III $\lambda$1123 panel as Fe III $\lambda$1123 at z = 0.166216 because there are no other related absorbers at this redshift including H $\lambda\lambda$1216, 1025, C II, C III, N II, N III, Si II, Si III, Si IV and Fe II,. We believe this line with an observed wavelength of 1309.0 Å is more simply explained as H I $\lambda$1216 at z = 0.07677. It is not possible to confirm the identification because FUSE observations are too noisy at the expected wavelength of the associated H I $\lambda 1025$.

### 4.1 Previous Studies

The system at $z = 0.16716$ has been studied by Chen & Prochaska (2000) and Prochaska et al. (2004). The system has a very strong multi-component Lyman series absorption from Ly $\alpha$ to Ly $\lambda$ 914. The relatively high H I column density in the system of log N(H I) = $16.45\pm0.05$ from Prochaska et al. (2004) is derived from a fit to the Lyman limit absorption.

Tripp et al. (2008) and Thom & Chen (2008) have also studied the absorber as part of their surveys of O VI in the IGM. Detailed profile fits to most of the H I and metal line absorbers are displayed and discussed in Thom & Chen (2008; see Fig. 22 and Table 15). The kinematical complexity of the $z = 0.16716$ absorber is evident from the multi-component fit to the Lyman series absorption of Lehner et al. (2006). Their fit included 5 H I components spanning a column density range of 3 dex and a velocity range of 141 km s$^{-1}$ in the rest frame of the absorber (see Table 9 and comment 6 in Lehner et al. 2006). The components from strongest to weakest were found to have redshifts of 0.16714, 0.16712, 0.16694, 0.16678 and 0.16661 corresponding to velocities of -5, -10, -57, -98, -141 km s$^{-1}$ when referenced to $z = 0.16716$. The component at v = -98 km s$^{-1}$ has a large but uncertain b = $74.9\pm7.2$: km s$^{-1}$ and if mostly thermally broadened may be associated with gas at T ~ $3.4\times10^5$ K which is near the temperature where O VI peaks in abundance under conditions of collisional ionization equilibrium. The three strongest H I components at v = -5, -10 and -57 km s$^{-1}$ have line widths consistent with origins in thermally and turbulently broadened photoionized gas with T ~ (1-3)x10$^4$ K. Other authors have proposed somewhat different component fits to the complex H I absorption (Prochaska et al. 2004; Thom & Chen 2008). This illustrates the ambiguities associated with component fitting to overlapping absorption lines in complex absorption systems. The COS data shows that the H I absorption actually extends over a wider velocity range with absorption in the wings of Ly $\alpha$ and $\beta$ detectable from -210 to +100 km s$^{-1}$.

No single one slab ionization model can be made to match the full range of observed ionization types seen in the absorber. Prochaska et al. (2004) demonstrated that the low and intermediate ions are approximately described by gas photoionized by the extragalactic background radiation with log U = -2.9±0.2 dex and a metallicity determined from Si to be [Si/H] = -0.25±0.2 dex. The background was taken to be the quasar only model of Haardt & Madau (2001). The modeling of the absorber is hampered by the fact that most of the important metal lines for constraining the model are saturated and therefore only provide lower limits to the column density. Prochaska et al. suggested the observations may imply an overabundance of N with respect to Si with [N/Si] > ~ 0.2 dex. They also suggested that O and C could have similar enhancements. Prochaska et al (2004) did not discuss in detail the physical condition implications of their photoionization model or the various possible origins of the highly ionized species.

*4.2 COS Observations of the z = 0.16716 system*

The COS observations allow us to evaluate the conditions in the absorber with measurements having much higher S/N than the STIS observation but at ~2.5 times lower resolution. The higher S/N allows us to search for weaker unsaturated absorbers and to evaluate the absorption in the velocity ranges where the absorption is weak and not detectable in the existing STIS measurements. Our goal here is not to perform a full analysis of the photoionized gas in the absorber as that would involve extensive multi-component profile modeling which is a major project because of the overlapping nature of the absorption and the ambiguities of performing profile fitting. We instead wish to use the previous and new observations to set approximate constraints on the physical conditions in the photoionized gas of the absorber. We then use those results to better understand the origin of the collisionally ionized gas in the absorber.

Table 2 lists measurements of the restframe equivalent widths, $W_r$, average absorption velocity, v, Doppler parameter, $b_a$, logarithm of the apparent column density, log $N_a$, and the integration range of the measurements. The measurements extend over the full range of the observed multiple component absorption. The apparent optical depth method of Savage & Sembach (1991) was used to determine $v_a$, $b_a$ and log $N_a$. The listed values of $b_a$ therefore do not allow for instrumental broadening but only provide an approximate indication of the observed line breadth. The column densities are listed as lower limits when there is strong evidence that the absorption contains unresolved saturation based on an inspection of the higher resolution STIS measurements and from the comparisons of $N_a(v)$ profiles for absorption lines of the same ion with a range of line strengths. The column densities we report use f-values from Morton (2003).

Unfortunately most of the low and intermediate ionization lines including C II, C III, N III, and Si III are strongly saturated and only lower limits to log $N_a$ can be obtained. The N II $\lambda1084$ line is mildly saturated. Plots of the $N_a(v)$ profiles for the Si II $\lambda\lambda1304, 1190, 1193$, and 1260 lines shown in the upper panel of Figure 4 reveal increasing line saturation from $\lambda1304$ to 1260. The measured values for the integrated column density for Si II $\lambda\lambda1190, 1193$ are log $N_a$ = 13.36±0.02 and 13.34±0.02, respectively. Since the stronger and weaker line differ in f-value by a factor of 2 we can use the saturation correction scheme of Savage & Sembach (1991) to derive a total saturation corrected Si II column density of log N = 13.38±0.03. This is consistent with log N(Si II) = 13.40±0.06 derived from the weakest Si II line and we adopt the value 13.38±0.03 for the Si II column density. This column density is 0.08 dex larger than the adopted column density of log N = 13.30±0.04 reported by Prochaska et al. (2004). Although their measurement of log $N_a$ = 13.42±0.05 for the Si II $\lambda1193$ line appears to be valid, we believe they made an error in reporting log $N_a$= 13.15±0.09 and a small restframe equivalent width for the Si II $\lambda1190$ line. The result for Si II is important because it provides one of the few reliable column densities for moderately ionized gas in the absorber.

The N II $\lambda1084$ line is moderately saturated. From the COS observations we obtain a lower limit to the column density $\log_a$N(N II) > 14.11 from the apparent column density. From the higher resolution STIS observations, Prochaska et al. (2004) report $\log_a$ N(N II) > 14.24. The COS N II absorption has the same maximum apparent optical depth as that for Si II $\lambda1260$ (see Figure 2). The Si II absorption requires a saturation correction of 0.21 dex to bring its apparent column density into agreement with the saturation corrected Si II column density derived from the weaker Si II $\lambda\lambda$ 1190, 1193 lines discussed above. If we adopt the same saturation correction for the N II $\lambda$1084 line we obtain log N(N II) = 14.32±0.10, with the estimated error mostly reflecting the uncertainty of the saturation correction.

The high S/N COS observations permits the detection of unsaturated Fe III $\lambda$1123 with log $N_a$ = 13.02±0.17. Fe II $\lambda$1145 is not detected with a 3$\sigma$ limit of log $N_a$ < 13.20. However, Fe II was detected in the much stronger $\lambda$2382 line by the FOS (Januzzi et al. 1998).

Prochaska et al. (2004) reported the possible detection of O I $\lambda\lambda$ 989, 1302 with $w_r$ = 20±6 mÅ and 30±10 mÅ, respectively. The COS spectrum reveals weak O I $\lambda\lambda$ 989, 1302 absorption with $w_r$= 7±4 and 13±3 mÅ, respectively. We claim a detection of the O I $\lambda$1302 line with log $N_a$(O I) = 13.24±0.09. O I and H I generally co-exist in the same gas because they are coupled by a strong charge exchange reaction. With log N(H I) = 16.45±0.04 we obtain log ( N(H I)/N(O I)) = 3.21±0.10 implying [O/H] = 0.10±0.10 using the photospheric solar oxygen abundance from Asplund et al. (2009) of (O/H) = $10^{-3.31}$.

The final adopted integrated column densities or limits for the multi-phase absorber are listed in Table 3 based on the measurements from COS, STIS and FOS in the case of C IV. The COS measurements are taken from Table 2 while the STIS observations are taken from Prochaska et al. (2004) except as noted.

$N_a(v)$ profiles for O VI $\lambda\lambda$ 1032, 1038 are shown in Figure 5. Except for v < -325 km s$^{-1}$ where C II $\lambda$1036 blends with the O VI 1038 absorption the two $N_a(v)$ profiles for O VI are in agreement, implying that the O VI absorption does not contain unresolved saturation.  The principal O VI absorption from +100 to -200 km s$^{-1}$ is strongly asymmetric and is clearly composed of multiple absorbing components with the strongest absorption near 0 km s$^{-1}$ and a weaker component near -135 km s$^{-1}$.

The relationship between the O VI absorption and absorption by lower stages of ionization is shown in Figure 6 which compares $N_a(v)$ profiles for O VI λ1032 with profiles for H I λ1216, N V λ1243, C III λ977, Si III λ1206, and Si II λ1190. The profiles for H I, C III and Si III have strong unresolved saturation near 0 km s$^{-1}$ and therefore do not properly trace the behavior of N(v) where the absorption is strongest. The absorption by O VI and N V is much broader than for the more weakly ionized gas. Although broader, the highly ionized gas is kinematically closely coupled to the more weakly ionized gas.  The O VI λ absorption in the broad component seen in Figure 3 extending from -200 to -400 km s$^{-1}$ with no associated absorption in other ions including H I is discussed in §5.

*4.3 Photoionization of the Moderately Ionized Gas*

In order to determine the conditions in the weakly and moderately ionized gas in the absorber we repeat the photoionization modeling discussed by Prochaska et al. (2004) but adopt the revised column densities from Table 3.

Photoionization calculations were performed using Cloudy [ver.C08.00, Ferland et al. 1998] and assuming a uniform density one zone slab model illuminated by the extragalactic background radiation. We use the Haardt & Madau (2001) extragalactic background radiation field incorporating photons from AGNs and star forming galaxies but adjusted to that appropriate for the redshift, z = 0.16716 of the absorption system.

We did not include possible ionizing radiation from the spiral galaxy in our approximate estimate of the background radiation field. For a roughly similar galaxy luminosity and impact parameter involving an absorber toward H 1821+643 at z = 0.225, Narayanan et al. (2010) found that the galaxy contribution to the ionizing background was ~ 15% of the Haardt & Madau (2001) background and therefore relatively unimportant. The same should be true for the LLS at z = 0.16716 toward PKS 0405-123. Note that the uncertainties associated with the adopted extragalactic background radiation field are larger than the additional contributions to the ionizing radiation from the galaxy.

The heavy element reference abundances assumed in the photoionization model are for the solar photosphere from the new abundance compilation of Asplund et al. (2009) and reflect the very large recent changes in the solar abundances of N, C , O , and Ne.

The photoionization modeling results are shown in Figure 7 for a model with log N(HI) = 16.45 , [O/H] = 0.10, [N/H] = 0.40 and [C/H] = 0, [Si/H] = -0.40 and [Fe/H] = 0. The different curves show how the expected ionic column densities change with log U where U = $n_\gamma / n_H$ is the ratio of ionizing photon density to total hydrogen (neutral +ionized) density. The heavy solid lines on the curves for each ion indicate the range of log U where the predicted column density agrees with the observed column density or its limit.

The absorber is extremely complex and the simple single-phase absorber assumed in the photoionization modeling only provides a very approximate assessment of the conditions in the absorber with multiple absorption components overlapping in velocity space.

Turning to species one might expect to be produced by photoionization in gas with T ~ (1-3)x10$^4$ K including H I, C II, C III, N II, N III, Si II, Si III, Fe II, Fe III, we find that the measured column densities of N II, Si II, Fe II, Fe III and the limits for C II, C III, and Si III are approximately consistent with the PI modeling for log U in the range from -3.0 to -3.8.

A deficiency of the model is it predicts substantially more O II than the observed upper limit of log N(O II) < 13.98 assuming b(O II) = 5 km s$^{-1}$. The absence of strong O II absorption is mystery we are unable resolve without access to higher quality measurements of the O II λ834 line which is redshifted to 974 Å. The model with [O/H] = 0.1 is consistent with the observed O I column density for log U <-3.4.

The uncertain abundances implied by the photoionization modeling are [O/H] = 0.1±0.10, [N/H] = 0.40±0.20, [Si/H] = -0.40±0.20 and [Fe/H] = 0 ±0.20. The fact that Fe requires log U ~ - 3.8 while Si requires log U ~ -3.1 reveals the inadequacy of the modeling. Therefore error estimates on these abundances are very uncertain. Changes in the adopted shape of the ionizing radiation field are known to substantially affect abundances derived from the simple photoionization modeling performed here (Howk et al 2008).

The major result we want to derive from the photoionization modeling is very approximate information about the physical conditions in the photoionized gas. For log U ~ -3.1 we infer log T ~ 3.9, n(H) ~ 10$^{-2.7}$ cm$^{-3}$, L ~ 1 kpc, log N(H) ~ 18.9, N(H I)/N(H) ~ 3.5x10$^{-3}$ and P/k ~ 40 cm$^{-3}$ K. The errors on these derived quantities are large given the simplicity of the photoionization modeling and the fact that the different elements require somewhat different values of log U. However, even with these uncertainties, it is clear the photoionized gas in this absorber is relatively cool and relatively confined.

An inspection of the figure immediately reveals that the absorption by O VI, N V and C IV can't be produced by photoionization in the gas producing the H I and weakly ionized metal lines. The expected column densities of O VI, N V and C IV in this cool photoionized gas with log U ~ -3.1 are 5.3, 2.9, and 1.7 dex smaller than the observed columns. The model's inability to produce the large observed column densities for the high ions is not surprising because the line profiles for the O VI and N V are much broader than for the low and intermediate ion profiles implying the low, intermediate, and high ions are not produced in the same gas.

If the O VI and N V ions are to be explained by photoionization, they require a medium with ionization parameters log U > -1.7 and > -2.0 which are ~ 1.2 dex larger than required for the weakly and moderately ionized gas. This in turn requires the path length, L, to be ~200 times larger, and $n_H$ and P to be ~10 times smaller than for the moderately ionized gas. While such a situation can't be ruled out, it is difficult to understand how the highly ionized gas could extend over such a large distance, be at such a low pressure, and be so well kinematically coupled to the weakly ionized gas as seen in Figures 3 and 6. A pure photoionization origin for most of the O VI and N V found in this absorber is unlikely.

*4.4 Collisional Ionization of the Highly Ionized Species*

The close kinematical association of the high ions to the low ion absorption over the velocity range from +50 to -50 km s$^{-1}$ is strongly suggestive of an origin of the O VI and N V in the interface regions of the cool photoionized gas and a hot ($T > 10^6$ K) exterior medium as found for the HVCs in the Milky Way (Sembach et al. 2003), the disk gas of the Milky Way (Bowen et al. 2009) and in the local interstellar medium (Savage & Lehner 2006). For the Milky Way HVCs, the column density of O VI is observed to range from log N(O VI) = 13.0 to 14.6. With log N(O VI) = 14.72±0.01, the absorber at z = 0.16716 has somewhat more O VI than found in the Milky Way HVCs. The O VI column density in the absorber is 0.23 dex smaller than the largest O VI column density so far measured in the low z IGM of log N(O VI) = 14.95±0.05 in a LLS at z = 0.20258 toward PKS 0312-77 (Lehner et al. 2009).

The observed values of the high ion column density ratios in the absorber are N(O VI)/N( V) = 6.8±0.03 and N(O VI)/N(C IV) = 3.6±0.4. The O VI to C IV column density ratio is similar to that found in Milky Way HVCs (see Table 11 of Sembach et al. 2003). However, the O VI to N V column density ratio is smaller than the lower limits found for the Milky Way indicating an enhanced amount of N V compared to O VI. This may be the result of the ~0.3 dex abundance enhancement of N compared to O seen in the photoionized absorber. If highly ionized interface absorbers are created in the evaporative phase of the interface, the abundances in the interface should reflect the abundances of the cooler medium being evaporated as a result of the heat transfer from the hot exterior medium.

The gas in the LLS is more highly ionized over the velocity range from -100 to -200 km s$^{-1}$ where absorption is only seen in H I, O VI and C III (see Figs. 3 and 6). Based on the $N_a(v)$ curves in Figure 6 we see that over the velocity range from -100 to -150 km s$^{-1}$ N(C III)/N(H I) ~ 0.5, N(O VI)/N(H I) ~2.5, and N(C III)/N(O VI) ranges from ~0.14 to 0.24. In highly ionized Galactic HVCs N(C III)/N(O VI) has been observed to range from <0.2 to >3.2 (Fox et al. 2006). When both ions yield reliable column densities, the range is from 0.3 to 1.5. The Galactic measurements are consistent with CIE in gas with log T ~ 5.2 or somewhat cooler gas in a non-equilibrium state. The behavior of C III and O VI in the LLS over the velocity range from -100 to -150 km s$^{-1}$ is similar to that found in Galactic highly ionized HVCs.

## 5. PROPERTIES AND IONIZATION OF THE O VI IN THE H I-FREE ABSORBER

*5.1 COS Observations of the H I Free Absorber*

We detect broad absorption from -190 to -350 km s$^{-1}$ in the O VI $\lambda$1032 line. The absorption also is evident in the O VI $\lambda$1038 line but that absorption blends with C II $\lambda$1036 in the z = 0.16716 system. The continuum-normalized profiles are shown in Figure 3. Figure 5 shows $N_a(v)$ profiles for both O VI absorbers overplotted. The rise in the value of $N_a(v)$ for O VI $\lambda$1037 compared to $\lambda$1032 for v < -325 km s$^{-1}$ is from blending with C II $\lambda$1036 absorption in the LLS.

The feature we identify as O VI $\lambda$1032 can't be caused by H I $\lambda\lambda$1025, 972, or higher Lyman series absorption because any associated stronger Lyman series lines are not detected at longer wavelengths. The feature is also not at a wavelength where it might be a metal line associated with any of the other known absorption systems seen in the spectrum of PKS 0405-123. Redshifted N IV $\lambda$765.15 in an associated absorption line system at z = 0.5726 would have a wavelength of 1203.3Å. However, the QSO does not have an associated system (Januzzi et al. 1998; Williger et al. 2006). The only alternate metal line identification we could find is S IV $\lambda$1062.66 associated with a Ly $\alpha$ absorber at z = 0.3233 with $W_r$ =142.5±13.5 mÅ and log N(H I) = 13.64±0.03 (Lehner et al. 2007). However, this identification is highly unlikely as S IV is rarely seen in metal-line systems and other metal lines that normally would be much stronger than S IV are not detected at z = 0.3233. Finally, the identification of the line at 1203.3 Å as O VI λ1032 is supported by the detection of the matching O VI λ1038 member of the doublet.

Measurements for the absorber are given in Table 4 using several measuring techniques. In the first row for O VI λ1032 we use the AOD method to produce the listed values of $v_a$, $b_a$, and log $N_a$(O VI) with the AOD integration extending from -190 to -400 km s$^{-1}$. The value of $b_a$ is not corrected for the effects of instrumental blurring of the absorption. In order to determine better parameters for O VI in the absorber we have employed the Voigt profile fitting routine developed by Fitzpatrick & Spitzer (1994). The COS line spread function used in the profile fit is that recommended by Ghavamian et al. (2009) for 1200Å. The profile fit was determined by seeking the minimum value of $\chi^2$ over the different velocity intervals from -400 to -190 km s$^{-1}$. We adopt the resulting values v = -278 ±3 km s$^{-1}$, b = 52 ±2 km s$^{-1}$, and log N(O VI) = 13.90±0.03 for O VI in the absorber. The reduced value of $\chi^2$ is 0.37 indicating an excellent fit to the assumed Voigt profile employed in the Fitzpatrick & Spitzer (1994) software. The errors on the absorber properties are the formal fit statistical errors but also include the small fit differences associated with performing the fit over the somewhat different velocity ranges of -450 to -210 km s$^{-1}$ and -400 to -210 km s$^{-1}$. The fitted profile for the preferred case is shown on the left panel of Figure 8. The actual velocity error is more like ±5 km s$^{-1}$ (absolute) when we include the COS wavelength calibration uncertainties discussed in §2. The profile fit yields an O VI column density 0.07 dex larger than that obtained from the apparent optical depth integration. The profile fit process process directly allows for instrumental blurring and therefore provides a better allowance for line saturation than the direct apparent optical depth integration provided the assumed line shape (Voigt profile) is correct.

The broad H I-free O VI $\lambda$1032 absorption is also visible in the much lower S/N STIS observations (see the feature near 1203.3 Å in the STIS spectrum shown in Figure 1 and the right panel of Figure 8 which shows velocity profile plots of the STIS measurements for both O VI lines binned to 10 km s$^{-1}$. The 1203.3 Å feature was not reported in any of the previous papers reporting STIS observations of PKS 0405-123 because its significance is low. We measure a restframe equivalent width of 71 ±24 mÅ over the velocity range from -200 to -368 km s$^{-1}$. The feature has less than 3 $\sigma$ significance. However, the right panel of Figure 8 shows an absorption feature exists in the STIS observation that agrees in velocity with the broad O VI line recorded by COS. Because of the very low S/N of ~ 5 per 7 km s$^{-1}$ STIS resolution element and low detection significance for the absorber, the results of a formal profile fit to the STIS data are not trustworthy. Therefore, we instead display the fit based on the COS results on the STIS observations.

The H I-free O VI absorber is of very great interest because it is very broad and symmetrical and is well fitted by an exp (-Gaussian) absorption profile convolved with the COS line spread function. The absorption must be tracing very highly ionized gas because there is no evidence for absorption by other species including H I $\lambda\lambda$1216, 1025, C II $\lambda$1036, N II $\lambda$1084. Si II $\lambda\lambda$1190, 1193, C III $\lambda$977, Si IV $\lambda\lambda$1394, 1404.

The -190 to -400 km s$^{-1}$ velocity region for possible absorption by Si III λ1206 is unfortunately contaminated by O VI 1038 at z = 0.36335 and Fe III λ1123 is contaiminated by H I λ1216 at z = 0.07677 (see the line contamination discussion at the beginning of §4).

There is evidence for weak N V λ1238 absorption over the -390 to -240 km s$^{-1}$ velocity range (see Figure 2). However, the derived strength is sensitive to the continuum placement because of the breadth and weakness of the line. The continuum in the vicinity of this apparent absorption exhibits undulations. For example, there is an unexplained ~10% elevation in the continuum from -400 to -600 km s$^{-1}$ (not displayed in Figure 2). The absorption is also affected by its overlap in the velocity range from -200 to -240 km s$^{-1}$ with the contaminating absorber 3 centered at – 270 km s$^{-1}$ which is H I λ1216 at z = 0.19255. Using the combined set of observations displayed in Figure 2 and our best estimate of the continuum level, we obtain for N V λ1238 over the velocity range from -240 to -390 km s$^{-1}$, $W_r$ = 33±4 mÅ, $v_a$ = -281±9 km$^{-1}$, $b_a$ = 51±12 km s$^{-1}$ and log $N_a$(N V) = 13.23±0.05. The errors here do not adequately allow for the continuum placement uncertainties. Also, for features this weak it is also important to carefully determine the impact of detector fixed pattern noise. By aligning the individual G160M integrations in the vicinity of the N V λ1238 absorption near 1443 Å in detector space we found that the G160M $\lambda_o$ = 1623 integration has a detector blemish near 1444.4 Å while a wire shadow strongly impacts the the G160M $\lambda_o$=1589 integration. Using only the observations obtained from the G160M $\lambda_o$= 1600 and 1611 integrations we obtain from -240 to -370 km s$^{-1}$ $W_r$ = 29±6 mÅ, v = -281±8 km$^{-1}$, $b_a$ = 45±6 km s$^{-1}$ and log $N_a$(N V) = 13.16±0.08. We prefer this second estimate of the parameters from the N V λ 1238 absorption since the observations seriously impacted by the fixed pattern noise structures have been omitted. The feature appears to be real but the error for the column density is ~50% larger than ±0.08 dex listed above because of the continuum placement uncertainty. We therefore adopt a N V column density of log N(N V) = 13.16±0.12.

The N V λ1242 feature is not detected with > 3σ significance. From the full set of combined observations displayed in Figure 2 we obtain when integrating from -220 to -370 km s$^{-1}$, $W_r$ = 4± 5 mÅ using our best estimate of the continuum. For a higher but acceptable continuum we obtain $W_r$ = 14±5 mÅ. This feature does not appear to be seriously affected by fixed pattern noise structures. The give the result for the λ1242 line as $W_r$ < 20 mÅ and log N(N V) < 13.27 in Table 4. If the feature revealed with the higher continuum placement for this line is real, it is consistent with the line strength of $W_r$ = 29±6 mÅ in the stronger member of the N V doublet. We conclude that N V associated with the H-free O VI absorption is probably detected with log $N_a$(N V) = 13.16±0.12. This implies log [N(O VI)/N(N V)] = 0.74±0.12 in the H-free O VI absorber. This is similar to the value log [N(O VI)/N(N V)] = 0.83±0.02 found in the much stronger LLS absorber over the 100 to -200 km s$^{-1}$ velocity range.

The absence of H I is important. The 3σ upper limits to the H I column density are given in Table 4 for different assumed values of b(H I) = 60, 100 and 150 km s$^{-1}$. The limits are derived from Voigt profiles overplotted on the Ly α line at the expected -278 km s$^{-1}$ velocity of H I absorption possibly associated with the broad O VI absorption. In Figure 9 we show the upper 50% of the normalized flux for the ±1000 km s$^{-1}$ region for H I Lyα in the z= 0.16716 reference frame. The large velocity interval allows an assessment of the behavior of the continuum well away from the position of possible H I absorption centered on -278 km s$^{-1}$. The irregular behavior of the continuum for v = 200 to 1000 km s$^{-1}$ is produced by some combination of weak IGM absorption and incomplete averaging of the detector fixed pattern noise when combining the individual integrations. The upper panel of Figure 9 displays Voigt profiles for b = 60 km s$^{-1}$ and log N(H I) = 12.7, 13.0, and 13.3 centered on v =-278 km s$^{-1}$ convolved with the COS LSF. The lower panel displays Voigt profiles for b = 150 km s$^{-1}$ and log N(H I) = 13.1, 13.4, and 13.7. Note that the spectral region near the center of possible broad H I absorption at -278 km s$^{-1}$ exhibits a weak emission feature with $w_r$ = -8±3 mÅ extending from -280 to – 320 km s$^{-1}$. The apparent emission feature with less than 3$\sigma$ significance hampers our search for broad H I Ly $\alpha$ absorption centered on v =-278 km s$^{-1}$. From the Voigt profiles illustrated in Figure 8 we estimate the 3$\sigma$ upper limits for H I associated with the broad O VI absorber to be log N(H I) < 12.7 for b = 60 km s$^{-1}$ and < 13.1 for b = 150 km s$^{-1}$. For b = 100 km s$^{-1}$ the limit is log N(H I) < 12.9.

The H I limit for b = 60 km s$^{-1}$ is valuable for evaluating the possibility that the O VI arises in a very low density highly photoionized medium with T ~ 5x10$^4$ K in which case the H I would be somewhat broader than the O VI if both species arise in the same gas. With log N(H I) < 12.7 , the implied H I to OVI column density ratio for the broad O VI absorber is N(H I)/ N(O VI) < 0.06. This is smaller than previously found for O VI absorbers in the intervening IGM. Other intervening absorbers with small values of N(H I)/N(O VI) include: (1) The z = 0.32639 system toward HE 0226-4110 studied by Thom & Chen (2008) with log N(O VI) = 13.6±0.2 and log N(H I) < 12.5 implying N(H I)/N(O VI) < 0.079. However, the O VI $\lambda$1038 line in this absorber is only detected with

~2σ significance so the actual origin of the line identified as O VI λ1032 is not secure. (2) The O VI absorber at z = 0.22638 in the spectrum of H 1821+643 studied by Narayanan et al. (2010) has log N(O VI) = 13.51±0.01 and log N(H I) <12.7 (3σ) implying N(H I)/N(O VI) < 0.15. (3) Tripp et al. (2008) reported log N(O VI) = 13.67±0.12 and log N(H I) = 12.72±0.12 for the absorber toward PG 1116+216 at z = 0.05927 implying N(H I)/N(O VI) = 0.11. This is the smallest ratio reported by Tripp et al. (2008) except for O VI in proximate systems where H I Ly α is often not seen. For the proximate systems, the small values of N(H I)/N(O VI) are likely the result of photoionization in the intense radiation environment near the AGN.

### 5.2. Ionization of the Broad H I-free O VI

We consider several possible origins for the ionization in the broad H I-free OVI absorber and conclude the O VI is most likely collisionally ionized in hot gas.

#### 5.2.1 A Photoionized Origin for the Broad H I-free O VI?

We first consider the possibility the absorber is produced by photoionization. Figure 10 shows the results of a model calculation assuming the O VI is produced by photoionization from the extragalactic background in a low density medium. The photoionization modeling assumptions and procedures are the same as for the study of the Lyman limit system (see §4.3). The model sets the H I column density at the observed upper limit log N(H I)< 12.7 and adjusts the oxygen abundance in order to achieve the observed column density of log N(O VI) = 13.90±0.03. The required model parameters with the minimum oxygen abundance of [O/H] = 0.0 are log U = -0.2, log

n(H) = -5.5, L = 130 kpc, log N(H)= 18.3, log T = 4.72, and P/k = 0.40 cm$^{-3}$ K. The required oxygen abundance would be larger if the true H I column density is less than the 3σ limit of log N(HI) <12.7.

The photoionization model which assumes a solar ratio of O/N predicts a N V column density of log N(N V) = 12.7 at log U = -0.2. This is 0.46 dex smaller than the observed column density of log N(N V) = 13.16±0.12 and suggests a supersolar value of [N/O] of 0.46 dex. The gas in the H I-free absorber may be related to the gas in the LLS for which we measure [N/O] = 0.3±0.2 (see §4.3)

If the absorber is produced by photoionization, it is necessary to explain the broad Gaussian-like shape of the absorber. The photoionized gas is relatively hot (5.2x10$^4$ K) due to photoionization heating with the large value of log U but the thermal Doppler parameter for the O VI at this temperature is $b_t$ = 7.4 km s$^{-1}$ while the observed value is b= 52±2 km s$^{-1}$. For the photoionization model to be valid, essentially all the line broadening would need to come from non-thermal origins such as turbulence or velocity gradients associated with a systematic flow. Turbulence does not generally give highly symmetrical Gaussian-like profiles unless the line of sight averages over a very large number of turbulent regions. Neither rotational broadening nor a tidal stream is likely to generate a broad Gaussian profile. We believe there is a much simpler explanation for the absorber involving collisional ionization in hot gas with log T > 6.

### 5.2.2 Collisional Ionization of the Broad H I-free O VI

The absorber is likely tracing hot collisionally ionized gas in the vicinity of the foreground galaxies. In Figure 11 we plot ion column densities versus temperature for gas in collisional ionization equilibrium (CIE) with solar elemental abundances from Asplund et al. (2009) and log N(H) = 19. The CIE curves are from the calculations of Gnat & Sternberg (2007). The set of curves include H I, O VIII, O VII, O VI, Ne VIII, N V, C IV, Si IV, and C III. The absence of H I with log [N(H I)/N(O VI)] < -1.2 implies log T > 5.4 for solar abundances of oxygen. However, at log T= 5.4 the thermal Doppler parameter for O VI is only 16.2 km s$^{-1}$, while the observed Doppler parameter is 52 ±2 km s$^{-1}$ which implies an upper limit to the temperature of log T = 6.41 if the broadening is dominated by thermal Doppler broadening. A temperature near log T = 5.4 would require a non-thermal component to the O VI line broadening with $b_{nt}$ ~ 49 km s$^{-1}$ with a Gaussian line shape. While we can not rule out such origins, it appears more likely the broadening is mostly controlled by thermal Doppler broadening. If the thermal Doppler contributions to the observed line width are the same as the non-thermal contributions with $b_t = b_{nt} = 36.8$ km s$^{-1}$, then log T = 6.11. Table 5 lists these and other combinations for the possible thermal and non-thermal broadening of O VI along with estimates of log N(H$^+$) and log N(H I) assuming a solar abundance ratio for (O/H) and collisional ionization equilibrium (CIE) .

The CIE model with a solar abundance ratio for O, N and H predictes log [N(O VI)/N(N V)] = 1.2 dex while the observed value is 0.74±0.12 dex. If the uncertain value of the N V column density in the H I-free O VI is correct, this implies the absorber has a supersolar value of [N/O] of 0.46±0.12 dex, which is roughly similar to the value [N/O] = 0.3±0.2 dex found for the LLS. The two absorbers may share a common nucleosynthetic history.

CIE should be a relatively good assumption at the higher temperatures listed in the table because at IGM densities high temperature gas cools slowly. Also, at these high temperatures and a redshift of 0.167, the modifications from CIE introduced by including the effects of the extragalactic background ionizing radiation are relatively small. This is illustrated in Figure 12 where we show how log [N(O VI)/N(H I)] depends on log U in a hybrid photoionization + collisional ionization model with solar abundances and temperatures of $10^5$, $3.2 \times 10^5$ and $10^6$ K. The hybrid model was calculated using Cloudy including the ionizing background radiation but fixing the temperature in the model. For T = $10^6$ K the value of log [N(O VI)/N(H I)] only begins to decrease for log U > -1.5 which corresponds to $n_H < 10^{-4}$ cm$^{-3}$. At log U = -1 corresponding to $n_H \sim 2 \times 10^{-5}$ cm$^{-3}$, the ionizing background radiation causes the CIE value of log [N(O VI)/N(H I)] change by -0.13 dex.

If the thermal and non-thermal broadening in the absorber are similar with $b_t$ = $b_{nt}$ = 36.8 km s$^{-1}$, then log T = 6.11, log N(H$^+$) = 19.87, and log N(H I) = 13.10. At this temperature H I would have a thermal line width of 146 km s$^{-1}$ and a total line width of 151 km s$^{-1}$. The corresponding H I Lyα line would have a central depth at line center of 6.5%. Such a H I absorption line is at the 3σ limit for H I discussed in §5.1 so the model and observations are consistent. If the oxygen abundance in the absorber were [O/H] = 0.1 dex as seen in the LLS, the expected H I absorption column density would decrease by 0.1 dex. If the gas in the absorber has log T = 5.8 we see from Table 5 that log N(H$^+$) = 19.11, log N(H I) = 12.75 and b(H I) = 111 km s$^{-1}$. For a H I line of this width the 3σ column density upper limit is log N(H I) ~ 12.9 and the model is again consistent with the absence of broad H I absorption.

33A subsolar abundance of oxygen in the absorber would cause the expected H I column density to be larger than the observed upper limit. For example for [O/H] = -0.2 and log T= 6.11, log N(H I) = 13.3 which is clearly inconsistent with the limit of 13.1 for b(H I) = 150 km s$^{-1}$ established in the lower panel of Figure 9.

It is interesting that the column density of O VI seen in the H I- free absorber, log N(O VI) = 13.90±0.03, is similar to that expected for the O VI associated with the very large column density of O VII detected at z = 0 toward AGNs. The O VII absorption probably traces a combination hot gas in the 3 kpc thick disk of the galaxy and a highly extended hot corona surrounding the Milky Way. Fang et al. (2003) report log N(O VII) = 16. 26 (+0.04, -0.25) for the line of sight to 3C 273 while Yao et al. (2008) find log N(O VII) = 16.00 (+0.16, -0.18) toward Mrk 421. If these lines of sight are tracing gas with T ~ (1-3)x10$^6$ K, the associated O VI will have log N(O VI) ~ 13.5 to 13.8 and b$_t$ ranging from 32 to 60 km s$^{-1}$. In the case of the Milky Way, it is difficult to see the O VI associated with the O VII because it blends with very strong O VI absorption produced by transition temperature gas with T ~ (1-5)x10$^5$ K in the Galactic thick disk with a scale height of ~2-3 kpc (Savage et al. 2003; Savage & Wakker 2009). If ~50% of the O VII absorption at z = 0 occurs in a hot extended corona surrounding the Milky Way, it's not unreasonable to expect to see such coronae surrounding external galaxies via broad but weak O VI absorption. Indirect evidence for an extended hot corona around the Milky Way comes from the detection of O VI in HVCs including the Magellanic Stream (Sembach et al. 2003) and the comet-like (head-tail) appearances of some of the high velocity clouds (Bruns et al. 2000). However, we note that Yao et al.

(2008) believe that most of the zero redshift O VI observed toward AGNs occurs within the 3 kpc thick disk of the Milky Way.

## 6. PHYSICAL ORIGINS OF THE ABSORBERS

### 6.1. *The Lyman Limit System*

The strong multi-phase LLS at $z = 0.16716$ is probably a HVC in the hot halo of one of the two foreground galaxies. The absorber may arise in the halo of the luminous spiral foreground galaxy at $z = 0.1668 \pm 0.0003$ with $\rho = 116\ h_{70}^{-1}$ kpc and $L = 4.4L^*$. The galaxy has a line of sight velocity of $-93 \pm 77$ km s$^{-1}$ with respect to reference redshift for the absorber. The near-solar abundances in the absorber suggests that the absorber may trace highly processed ejecta from the galaxy. The ejecta may have been produced during an episode of star formation ~1 Gyr ago that would account for the current population of A stars seen in the integrated spectrum of the associated luminous spiral galaxy (Prochaska et al.2006). An alternate possibility is the HVC is tracing a tidal stream from a galaxy interacting with the hot halo of the spiral galaxy. It is possible that the dwarf galaxy near the luminous spiral galaxy has produced a tidal stream similar to the Magellanic Stream.

Gas in a hot halo surrounding the approximately face-on spiral galaxy would be expected to have a velocity roughly similar to the systemic velocity of the galaxy. Therefore, any O VI associated with the hot halo gas at $T > 10^6$ K would be expected to have its absorption blended with the LLS absorption. Perhaps some of the strong O VI absorption we have attributed to gas in the interfaces between the LLS and a hot exterior medium is in fact O VI absorption tracing the hot halo gas. If the gas in the hot halo is kinematically disturbed it is possible that the H I-free broad O VI absorber at -278 km s$^{-1}$ is tracing that hot halo (see the following section).

### 6.2. The Broad H I–free O VI Absorber

It's tempting to associate the broad H I-free O VI absorption with a hot gaseous corona surrounding the luminous foreground spiral galaxy. However, such an assignment may be inconsistent with the O VI absorption velocity of –278±3 km s$^{-1}$ compared to the galaxy velocity of -93±77 km s$^{-1}$ implying a velocity difference of -185± 77 km s$^{-1}$. For a face-on or moderately inclined galaxy one would expect the corona to absorb at approximately the same velocity as the galaxy. The origin of the velocity difference could be from an outflow of hot gas from one side of the spiral galaxy or from the dwarf galaxy. If the two galaxies share a kinematically-disturbed common halo, the observed velocity differences also would be easy to explain. A shock structure in the corona of one of the galaxies is possible, but with this explanation, it is difficult to understand the symmetric appearance of the absorber and the absence of associated lower temperature ions. With a single sight line though the environment of the foreground galaxies it is not easy to choose among the many possibilities. Numerical simulations of galaxy formation and feedback suggest that the hot gas structures in circum-galactic environments will be complex (Oppenheimer & Davé 2009). Although the galaxy impact parameters are relatively small ($\rho \sim 100$ kpc) and strongly suggest an association with one or both galaxies, we can't rule out the possibility that the absorbers trace a hot gas filament or a galaxy group in which the two galaxies are imbedded. Indeed, the velocity offset of the hot absorber relative to the galaxies plus the absorber's symmetric line profile suggest an equilibrium association with a structure centered at the absorber velocity.

Given our lack of knowledge of the geometrical shape, extent or location of the absorber, the associated mass estimates are extremely uncertain but interesting because the implied column densities of $H^+$ are so large. If the thermal and non-thermal broadening of the O VI are the same, log T = 6.11 and log $N(H^+)$ = 19.87. If this gas extends over a 200 kpc thick region, the average gas density is $n_{H+}$ = $1.2 \times 10^{-4}$ cm$^{-3}$ and the pressure P/k = 270 K cm$^{-3}$. A spherical, 200 kpc diameter region with an average density $1.2 \times 10^{-4}$ cm$^{-3}$ would have a mass of $\sim 1.6 \times 10^{10}$ M$_\odot$.

We speculated in §3 that the galaxy redshift survey in this region maybe somewhat incomplete so that it is possible that this hot absorber lies close to the physical center of a small group of late-type galaxies, the most abundant type of galaxy association in the Universe. The absence of lower ions suggests a hot, collisionally-ionized gas cloud. Further, the symmetrical line profile of this absorber suggests that the hot gas is in equilibrium with some larger dark matter structure, like a filament or galaxy group potential well. Mulchaey et al. (1996) hypothesized that if spiral-rich galaxy groups are gravitationally-bound entities, they might contain a gravitationally-heated reservoir of log T ~ 6.3 K hot gas due to their ~ 100 km s$^{-1}$ velocity dispersion. This gas would be too cool to emit sufficient thermal bremsstrahlung X-rays to be detectable by the X-ray telescopes of that decade but would be seen in absorption by high ions like O VI against background QSOs. But in this case, Mulchaey et al. (1996) note, lower ions like N V, C IV and Si IV would not be detected, although the total hydrogen column density would be quite high (log N(H) ~ 20 ) and almost completely ionized. This hypothesis was developed by extrapolation to lower group velocity dispersions based on the X-ray detections of groups of galaxies dominated by one or more, bright ellipticals (Mulchaey & Zabludoff 1998). The physical parameters (symmetrical line profile in O VI and absence of lower ion absorption) suggested for this scenario, including a total gas mass of $\sim 10^{10}$ M$_\odot$ (about an order of magnitude less gas than in elliptical-dominated groups), are quite similar to those derived for the hot PKS 0405-123 absorber reported here, although the observed O VI line width is somewhat narrower than expected for a virialized group. New galaxy survey data can confirm or deny the presence of a galaxy group at z=0.167 to support or refute this hypothesis.

If most spiral-rich galaxy groups contain $\sim 10^{10-11}$ M$_\odot$ of $\sim 10^6$ K hot gas (corresponding to a 200-400 kpc diameter sphere of gas with n $\sim 1.2 \times 10^{-4}$ cm$^{-3}$, this may be a significant total reservoir of WHIM gas in the local Universe not yet accounted for by the current baryon census. For example, using a cosmological volume density of 7 x $10^{-3 \text{ to } -4}$ Mpc$^{-3}$ for groups of galaxies with total L $\geq$ L$^*$ and/or velocity dispersion $\geq$ 100 km s$^{-1}$ from Girardi & Giuricin(2000) and Pisani, Ramella & Geller (2003), such hot gas spheres would constitute a baryon reservoir of 1.5-15% of the total $\Omega_b$ at low z. If instead these broad, shallow O VI-only absorbers are associated with gaseous filaments, then their physical sizes and cosmological frequencies are not well-constrained and so the only constraint on their baryon content would then come from detecting more examples of these systems. There is little chance that any similar system would have been detected here-to-fore because this first detection required the very high S/N obtained by COS on this very bright AGN target at a redshift low enough to find associated foreground galaxies. At lower redshifts where galaxy groups are easier to identify, the O VI doublet lies in the *FUSE* FUV spectral band. Quite generally, the S/N in *FUSE* spectra of AGN are too low to detect absorbers like this one. There is one exception to this generalization, the high-S/N *FUSE* spectrum of PKS 2155-304 contains a pair of weak O VI absorbers flanking high-N(H I) absorbers at the redshift of a spiral-rich foreground group of galaxies at cz ≈ 17,000 km s$^{-1}$ (Shull, Tumlinson & Giroux 2003). Neither C IV nor Si III is detected in the STIS spectrum reported in Shull et al., while Fang et al. (2002) reported a tentative O VIII absorption detection in PKS~2155-304 with *Chandra* at a redshift between the two O VI absorbers. This detection has not been confirmed by *XMM/Newton* observations (Rasmussen, Kahn & Paerels 2003), nor is it decidedly refuted. Shull et al. explore a clumpy infall and shock ionization model for these absorbers and arrive at similar physical properties for the intra-group medium in the 17,000 km s$^{-1}$ group as we have derived here for the z=0.166 system. Thus it seems worthwhile to use COS to search for other, similar systems. On the other hand, detecting diffuse hot gas in spiral-rich groups in either emission or absorption would require prohibitively long observations with *Chandra* or *XMM/Newton* (we estimate $\geq$ 1 Msec with *Chandra* to detect the hot group gas in the PKS~0405-123 sightline; however, see Rasmussen et al. 2006 for *XMM/Newton* detections of a few high-velocity dispersion, spiral-dominated groups. Even tentative X-ray absorption detections of O VII and/or O VIII will probably remain controversial until the next generation of X-ray observatories.

At higher redshifts ground-based spectroscopy can obtain exquisite high S/N spectra of numerous QSO targets. However, any broad, shallow O VI absorbers would lie within the Lyα forest and so would be very difficult to find. It is not surprising that the HI-free OVI absorber population has not been detected before now.

Ultraviolet measures with COS of O VI potentially provide a sensitive way of directly detecting gas in the IGM with T > $10^6$ K with a sensitivity and spectral resolution substantially higher than will be provided for O VII at X-ray energies for many years to come. This occurs because of the high temperature tail seen on the plot of log N(O VI) vs log T in Figure 11. This tail is caused by strong dielectronic recombination of O VII into O VI at log T ~ 5.8 to 6.2. Therefore, O VI absorption with 32 km s$^{-1}$ < $b_{obs}$ < 60 km s$^{-1}$ and little or no associated H I is potentially revealing hot gas with log T from ~5.8 to 6.2. The principal problem in using the O VI as a diagnostic of hot gas with log T > 6 will be determining the relative contributions of non-thermal and thermal broadening to the observed absorption line width and shape. If the weak but very broad associated H I absorption can be detected, it will be possible to solve for the thermal and non-thermal contributions to the broadening given the factor of 16 difference in mass between O VI and H I. Additional information about the temperature of the plasma can be provided if the absorption is at a redshift that allows the detection of Ne VIII $\lambda\lambda$770, 780 since for log T >5.75 the column density of Ne VIII should exceed that for O VI (see Fig. 11) assuming a solar ratio for O/Ne. Unfortunately, possible Ne VIII $\lambda$770, 780 in the H I-free z = 0.166 O VI system lies just short-ward of the ISM absorption cutoff at 912 Å. However, for other systems it may be possible to search for Ne VIII with COS provided z > 0.47.

The measurements of this paper demonstrate the utility of using O VI as a diagnostic of million degree gas near galaxies. Gas this hot was previously believed to be traceable only through higher ionization stages at X-ray wavelengths. Measures of O VI (and Ne VIII) along other sight lines will determine whether the H I-free O VI absorber is an isolated example, or whether such observations will revolutionize our understanding of hot ( $T > 10^6$ K ) gas in the vicinity of galaxies.

## 7. SUMMARY

1. High S/N observations of the QSO PKS 0405-123 ($z_{em}$ = 0.572) with the Cosmic Origins Spectrograph (COS) from 1134 to 1796 Å with a resolution of ~17 km s$^{-1}$ are used to study the properties of the multi-phase partial Lyman limit system at z = 0.16716 and an associated broad H I-free O VI absorber. The LLS has previously been studied by Prochaska et al. (2004) using relatively low S/N spectra from STIS and FUSE. The broad H I-free O VI absorber may represent a rare but important new class of low z IGM absorbers.

2. The LLS and the associated H I-free broad O VI absorber likely originate in the circumgalactic gas associated with a pair of galaxies at z = 0.1688 and 0.1670 with impact parameters of 116 $h_{70}^{-1}$ and 99 $h_{70}^{-1}$ and L = 4.4L$^*$ and 0.15L*, respectively.

3. The broad and symmetric O VI absorption is well characterized by absorption at v = -278±3 km s$^{-1}$ with log N(O VI) = 13.90±0.03 and b = 52±2 km s$^{-1}$. This broad absorption feature is not detected in H I λ1216 with a 3σ column density limit of log N(H I) < 12.7 if b(H I) ~ 60 km s$^{-1}$ or log N(H I) < 13.1 if b(H I)~ 150 km s$^{-1}$. The broad and symmetric appearance of the O VI profile suggests thermal broadening in a hot gas with log T ~6.1.

4. If the broad H I-free O VI absorber is tracing a spherical gaseous structure of with ~200 kpc diameter the mass is ~1.6x10$^{10}$ (O$_\odot$/O)M$_\odot$ for log T = 6.1.

5. The LLS absorber has strong asymmetrical O VI absorption spanning a velocity range from -200 to +100 km s$^{-1}$ in the z= 0.16716 rest-frame with log N(O VI) = 14.69±0.02. The new COS observations also provide high quality measures of H I, C III, C II, N V, N III, N II, Si IV, Si III, Si II, O I and Fe III.

5. Photoionization modeling of the low and intermediate ions of the multi-phase absorber provide approximate constraints for the physical conditions in the LLS. For log U ~ -3.1 we infer log T ~ 3.9, n(H) ~ 10$^{-2.7}$ cm$^{-3}$, L ~ 1 kpc, log N(H) ~ 18.9, N(H I)/N(H) ~ 3.5x10$^{-3}$ and P/k ~ 40 cm$^{-3}$ K. The photoionized gas in this absorber is relatively cool and relatively confined.

6. The kinematic relationships between the O VI absorption and the low ionization absorbers in the LLS are similar to those found in Milky Way HVCs including the tidal debris of the Magellanic Stream which suggests that much of the O VI is produced by collisional ionization in the interface regions between cooler photoionized gas and a hot exterior medium with T > 10$^6$ K.

7. High S/N COS observations of broad O VI absorbers like the one observed at z = 0.167 toward PKS 0405-123 allow the direct detection of hot (T >10$^6$ K) gas associated with galaxies.

Acknowledgements: We thank the NASA astronauts for a great job in successfully completing the large number of difficult tasks during the 2009 HST servicing mission. We thank the many people involved with building COS and with determining its flight performance characteristics. BDS and AN acknowledge funding support from NASA through the COS GTO contract to the University of Wisconsin-Madison through the University of Colorado. This work was supported by NASA grants NNX08-AC146 and NAS5-98043 to the University of Colorado at Boulder.

References

Asplund, M., Grevesse, N. Sauval, A. J., & Scott, P. 2009, ARAA, 47, 481

Bowen, D. V., Jenkins, E. B. et al. 2008, ApJS, 176, 59

Bregman, J. 2007, ARAA, 45, 221

Bruns, C., Kerp, J., Kalberla, P. M. W., & Mebold, U. 2000, A&A, 357, 120

Cen, R., & Ostriker, J. P. 1999, ApJ, 514, 1

Cen, R., & Fang, T. 2006, ApJ 650, 573

Chen, H.-W., & Mulchaey, J. S. 2009, ApJ, 701, 1219

Chen, H.-W., & Prochaska, J. X. 2000, ApJ, 543, L9

Chen, H-W., Prochaska, J.X., Weiner, B. J., Mulchaey, J.S. & Williger, G.M. 2005, ApJ, 629, L25.

Danforth, C. W., & Shull, J. M. 2008, ApJ, 679, 194

Danforth, C. W., Stocke, J. T., & Shull, J. M. 2010a, ApJ, 710, 613

Danforth, C.W., Keeney, B. A., Yao, Y., Stocke, J. T. & Shull, J. M. 2010b, ApJ, (in preparation).

Davé, R. Cen, R., Ostriker, J. P.; Bryan, G. L., Hernquist, L., Katz, N., Weinberg, D. H., Norman, M. L., & O'Shea, B. 2001, ApJ, 552, 473

Dixon, W. V. et al. 2010, Cosmic Origins Spectrograph Instrument Handbook, version 2.0 (Baltimore, STSCI)

Fang, T., & Bryan, G. L. 2001, ApJ, 561, L31

Fang, T., Marshall, H. L., Lee, J. C., Davis, D. S. & Canizares, C. R. 2002, ApJL, 572, L127

Fang, T., Sembach, K. R., & Canizares, C. R. 2003, ApJ, 586, L49

Ferland, G. J., Korista, K. T., Verner, D. A., Ferguson, J. W., Kingdon, J. B., & Verner, E. M. 1998, PASP, 110, 761

Fitzpatrick, E. L., & Spitzer, L. 1994, ApJ, 427, 232

Fox, A. J., Savage, B. D. & Wakker, B. P. 2006, ApJS, 165, 229

Froning, C., & Green, J. C. 2009, Ap&SS, 320, 181

Ghavamian, P. et al. 2009, COS Instrument Science Report, HST Science Institute, COS ISR 2009-01(v1)

Girardi, M., & Giuricin, G. 2000, ApJ, 540, 45

Gnat, O., & Sternberg, A. 2007, ApJS, 168, 213

Green, J. et al. 2010, ApJ (in preparation)

Haardt, F., & Madau, P. 2001, in Clusters of Galaxies and the High Redshift Universe Observed in X-rays, Recent results of XMM-Newton and Chandra, XXIst Moriond Astrophysics Meeting, March 10-17, 2001 Savoie, France, eds, D.M. Neumann & J.T.T. Van

Howk, J. C., Ribaudo, J. S., Lehner, N., Prochaska, J. X., & Chen, H.-W. 2009, MNRAS, 396, 1875

Januzzi, B. et al. 1998, ApJS, 118, 1

Lehner, N., Prochaska, J. X., Kobulnicky, H. A., Cooksey, K. L., Howk, J. C., Williger, G. M., Cales, S. L. 2009, ApJ, 694, 734

Lehner, N., Savage, B. D., Richter, P., Sembach, K. R., Tripp, T. M., & Wakker, B. P. 2007, ApJ 658, 680

Lehner, N., Savage, B. D., Wakker, B. P., Sembach, K. R., & Tripp, T. M. 2006, ApJS, 164, 1

Lockman, F. J., & Savage, B. D.1995, ApJS, 97, 1

Ménard, B., Wild, V., Nestor, D., Quider, A. & Zibetti, S. 2009, astro-ph/0912.3263

Morris, S., Weymann, R. J. et al. 1993, ApJ, 419, 524

Morton, D. C. 2003, ApJS, 149, 205

Mulchaey, J.S. Mushotzky, R.F., Burstein, D. & Davis, D.S. 1996, ApJ, 456, L8

Mulchaey, J.S. & Zabludoff, A. I. 1998, ApJ, 496, 73

Narayanan, A., Savage, B. D., & Wakker, B. P. 2010, ApJ, (submitted)

Narayanan, A., Wakker, B. P., & Savage, B. D. 2009, ApJ,703, 74

Oppenheimer, B., & Davé, R. 2008, MNRAS, 387, 577

__________________ 2009, MNRAS, 395, 1875

Osterman, S., et al. 2010, ApJ (in preparation)

Penton, S. V., Stocke, J. T., & Shull, J. M. 2002, ApJ, 565, 720

______________________________2004, ApJS, 152, 29

Pisani, A., Ramella, M., & Geller, M.J. 2003, AJ, 126, 1677

Prochaska, J. X., Chen, H.-W., Howk, J. C., Weiner, .J., & Mulchaey, J. S. 2004, ApJ, 617, 718

Prochaska, J. X., Weiner, B. J., Chen, H.-W, & Mulchaey, J. S. 2006, ApJ, 642, 989

Rasmussen, A., Kahn, S.M. & Paerels, F., 2003, in the IGM/Galaxy Connection, (Dordrecht: Kluwer Publ.), ed. J. Rosenberg & M. Putman, 109

Rasmussen, J., Ponman, T.J., Mulchaey, J.S., Miles, T.A., & Raychaudhury, S., 2006,

MNRAS, 373, 653

Richter, P., Savage, B. D., Tripp, T. M., & Sembach, K. R. 2006, A&A, 451, 767

Savage, B. D., Wakker, B. P., Sembach, K. R. et al. 2003, ApJS, 146, 125

Savage, B. D., Lehner, N., Wakker, B. P., Sembach, K. R., & Tripp, T. M. 2005, ApJ, 626, 776

Savage, B. D., & Sembach, K. R. 1991, ApJ, 379, 245

Savage, B. D., & Lehner, N. 2006, ApJS, 162, 134

Savage, B. D., & Wakker, B. P. 2009, ApJ, 702, 74

Sembach, K. R., Wakker, B. P., Savage, B. D. et al. 2003, ApJS, 146, 165

Shull, J.M., Tumlinson, J., & Giroux, M. L. 2003, ApJ, 594, L107

Spinrad, H. et al. 1993, AJ, 106, 1

Stocke, J., T., Penton, S. V., Danforth, C. W., Shull, J. M., Tumlinson, J., & McLin, K. M. 2006, ApJ, 641, 217

Thom, C., & Chen, H.-W. 2008, ApJS, 179, 37

Tripp, T. M., Lu, L., & Savage, B. D. 1998, ApJ, 508, 200

Tripp, T. M., Sembach, K. R., Bowen, D. V., Savage, B. D., Jenkins, E. B., Lehner, N., & Richter, P. 2008, ApJS, 177, 39

Wakker, B. P., & Savage, B. D. 2009, ApJS, 182, 378

Williger, G. M., Heap, S. R., Weymann, R. J., Davé, R., Ellingson, E., Carswell, R. F., Tripp, T. M., & Jenkins, E. B. 2006, ApJ, 630, 631

Yao, Y., Nowak, M. A., Wang, Q. D., Schulz, N. S., & Canizares, C. R. 2008, ApJ, 672, L21

46Table 1. COS G130M and G160M Integrations of PKS 405-123[a]

| MAST ID | Date[a] | Grating | $\lambda_c$ [Å] | $\lambda_{mim}$ [Å] | $\lambda_{max}$ [Å] | t [sec] |
|---|---|---|---|---|---|---|
| LACB51010 | 8/31/09 | G130M | 1291 | 1134 | 1430 | 1984 |
| LACB51020 | 8/31/09 | G130M | 1300 | 1146 | 1440 | 650 |
| LACB51030 | 8/31/09 | G130M | 1309 | 1156 | 1449 | 1908 |
| LACB51040 | 8/31/09 | G130M | 1300 | 1146 | 1440 | 650 |
| LACB51050 | 8/31/09 | G130M | 1318 | 1165 | 1458 | 1908 |
| LACB51060 | 8/31/09 | G130M | 1300 | 1146 | 1440 | 650 |
| LACB51070 | 8/31/09 | G130M | 1327 | 1175 | 1468 | 1908 |
| LB6822010 | 12/21/9 | G160M | 1589 | 1401 | 1761 | 2167 |
| LB6822020 | 12/21/9 | G160M | 1600 | 1412 | 1772 | 2965 |
| LB6822030 | 12/21/9 | G160M | 1611 | 1424 | 1784 | 2965 |
| LB6822040 | 12/21/9 | G160M | 1623 | 1436 | 1796 | 2965 |
| LB6823010 | 12/21/9 | G130M | 1291[b] | 1141 | 1427 | 4830 |

[a]The table lists the MAST exposure ID number, the date of observation, the COS grating, the central set-up wavelength, the minimum and maximum wavelengths covered by the integration, and the exposure time.

[b]These observations were obtained at PF-POS = 1, 2, 3 and 4 positions with integration times of 1000, 1400, 1000, and 1430 sec, respectively.

Table 2. COS Measurements for the Multi-phase Absorber at z = 0.16716[a]

| ion | $\lambda_r$ | $w_r$ (mÅ) | $v_a$ (km s$^{-1}$) | $b_a$ (km s$^{-1}$) | log $N_a$ (dex) | [-v, +v] (km s$^{-1}$) | Note |
|---|---|---|---|---|---|---|---|
| H I | 1216 | 900±6 | -23±3 | 90±2 | >14.8 | [-240, 190] | 1 |
| H I | 1025 | 518±8 | -12±5 | 64±3 | >15.0 | [-240, 190] | 1 |
| O I | 989 | 7±4 | -18±6 | 11±6 | <13.48 | [-40, 10] | 2 |
| O I | 1302 | 13±3 | -20±4 | 17±3 | 13.24±0.09 | [-40, 10] | |
| O VI | 1032 | 418±7 | -47±2 | 79±3 | 14.71±0.01 | [-210, 120] | |
| O VI | 1038 | 264±9 | -39±1 | 83±4 | 14.73±0.02 | [-210, 120] | |
| C II | 1036 | 162±6 | -13±1 | 39±3 | >14.32 | [-90, 85] | 1 |
| C II | 1335 | 219±5 | -20±2 | 37±4 | >14.22 | [-120, 55] | 1 |
| C III | 977 | 479±8 | -48±1 | 75±3 | >14.20 | [-210, 110] | 1 |
| C IV | 1548 | 452±39 | 0±50 | 129±4 | 14.11±0.04 | [-190, 180] | 3 |
| C IV | 1550 | 280±61 | 0±50 | 144±15 | 14.19±11 | [-190,180] | 3 |
| N II | 1084 | 104±4 | -13±1 | 28±6 | >14.11 | [-90, 85] | 1 |
| N III | 990 | 216±4 | -4±1 | 42±3 | >14.51 | [-90, 85] | 1,4 |
| N V | 1238 | 136±4 | -26±3 | 43±3 | 13.88±0.01 | [-90, 85] | |
| N V | 1243 | 77±6 | -23±2 | 44±4 | 13.91±0.03 | [-90, 85] | |
| Si II | 1190 | 71±4 | -15±2 | 34±7 | 13.36±0.02 | [-110, 70] | |
| Si II | 1193 | 125±4 | -12±3 | 39±3 | 13.34±0.02 | [-110, 70] | |
| Si II | 1260 | 163±4 | -18±2 | 38±4 | >13.17 | [-110, 70] | 1 |
| Si II | 1304 | 31±3 | -10±3 | 31±5 | 13.40±0.06 | [-70, 50] | |
| Si II | 1526 | 67±20 | -13±4 | 35±4 | 13.54±0.14 | [-70, 50] | |
| Si III | 1206 | 248±4 | -11±1 | 32±4 | >13.33 | [-90, 85] | 1 |
| Si IV | 1394 | 94±6 | -8±2 | 37±7 | 13.10±0.03 | [-75, 100] | |
| Si IV | 1403 | 70±5 | -7±4 | 41±8 | 13.21±0.04 | [-75, 100] | |
| Fe II | 1145 | 12±3 | … | 26±7 | <13.19 | [-20, 20] | 5 |
| Fe II | 2383 | 120±30 | 0±50 | 80±9 | 12.92±0.13 | [-100, +125] | 6 |
| Fe III | 1123 | 9±3 | -2±10 | 23±7 | 13.02±0.17 | [-32, 32] | |

[a] Measurements listed here are from COS with the exception of C IV λλ 1548, 1550 and Fe II λ2383 which are from FOS (Januzzi et al. 1998). Except where noted, all the column densities, velocities and line widths are derived from the AOD method. The AOD line widths, $b_a$, are not corrected for the effects of instrumental blurring.

Notes: (1) The absorption is strongly saturated. The measured value of log $N_a$ is reported as a lower limit. (2) The O I 989 line is not detected with >3σ significance. The upper limit was obtained from $w_r$ < 12 mÅ assuming the absorption line is on the linear part of the COG. (3) C IV measurements are from the FOS (Januzzi et al. 1998). We list the measured values of log Na. With b = 80 km s$^{-1}$ a COG solution yields log N(C IV) = 14.16± 0.05. That value is listed in Table 3. We attribute all the C IV absorption to gas in the LLS. (4) The N III measurement includes weak contamination from Si II λ990. (5) Although a feature of 4σ significance exists at the expected position of Fe II λ1145, the absorption line appears to be blended with another weak IGM absorber or is affected by fixed pattern noise. We therefore report a limit for Fe II 1145 using $w_r$ < 15 mÅ assuming no line saturation. (6) The Fe II λ2383 observation is from the FOS (Januzzi et al. 1998) but measured by us. The observed rest frame equivalent width implies log N(Fe II) = 12.87±0.09 if there is no line saturation. Including a

0.09 dex saturation correction based on the saturation effects observed for the Si II lines we obtain log N(Fe II) = 12.95±0.15 which is slightly larger than the value of Na listed above. Given the low resolution of the FOS observation we adopt log N(Fe II) = 12.95±0.15.

Table 3. Adopted Total Column Densities in the Multi-Phase LLS at z = 0.16716

| Ion    | log N        | Instrument | Comment |
|--------|--------------|------------|---------|
| H I    | 16.45±0.05   | STIS       | 1       |
| C II   | >14.32       | COS        | 2       |
| C III  | >14.20       | COS        | 2       |
| C IV   | 14.16±0.05   | FOS        | 3       |
| N II   | 14.32±010    | COS        | 4       |
| N III  | >14.51       | COS        | 2, 5    |
| N V    | 13.89±0.02   | COS        | 2       |
| O I    | 13.24±0.09   | COS        | 2       |
| O II   | <13.98       | STIS       | 6       |
| O VI   | 14.72±0.01   | COS        | 2       |
| Si II  | 13.38±0.03   | COS        | 2       |
| Si III | >13.33       | COS        | 2       |
| Si IV  | 13.15±0.05   | COS        | 2       |
| Fe II  | 12.95±0.15   | COS        | 5       |
| Fe III | 13.02±0.17   | COS        | 2       |

Comments: (1) We have adopted the STIS measurement of Prochaska et al. (2004). (2) Result is based on the COS results from this paper listed in Table 2. (3) Value is from the 230 km s$^{-1}$ resolution FOS observation discussed in note 3 to Table 2. (4) We adopt the saturation corrected N II column density discussed in §4.2. The correction is inferred from the saturation observed for the Si II $\lambda\lambda$1190, 1193, 1260 absorption. (5) This limit is affected by contamination from Si II $\lambda$990. (6) Prochaska et al. (2004) report for the O II $\lambda$834 line $w_r < 20$ mÅ and log N(O II) < 13.65. We repeated the measurement and obtain a 3$\sigma$ upper limit of $w_r < 42$ mÅ integrating over the velocity range from -40 to 10 km s$^{-1}$. For b = 10 km s$^{-1}$ log N(O II) < 13.58. For b = 5 km s$^{-1}$ the limit becomes <13.98. We adopt the larger limit.

Table 4. COS Measurements for the Broad H I-free Absorber[a]

| ion | $\lambda_r$ | $W_r$ (mÅ) | v (km s$^{-1}$) | b (km s$^{-1}$) | log N (dex) | [-v, +v] (km s$^{-1}$) | method |
|---|---|---|---|---|---|---|---|
| O VI | 1032 | 78±69 | -289±5 | 60±5 | 13.83±0.04 | [-400, -190] | AOD |
| O VI | 1032 | … | -278±3 | 52±2 | 13.90±0.03 | [-400, -190] | PF[b] |
| H I Ly α | 1216 | … | -278 | 60 | <12.7[c] | … | 3σ limit[c] |
| H I Ly α | 1216 | … | -278 | 100 | <12.9[c] | … | 3σ limit[c] |
| H I Ly α | 1216 | … | -278 | 150 | <13.1[c] | … | 3σ limit[c] |
| N V | 1238 | 29±6 | -281±8 | 45±6 | 13.16±0.12 | [-370, -240] | AOD[d] |
| N V | 1242 | <20 | … | … | <13.27 | [-370, -220] | 3σ limit[δ] |
| Si IV | 1394 | <24 | … | … | <12.43 | [-400, -190] | 3σ limit |
| Si IV | 1403 | <24 | … | … | <12.62 | [-400, -190] | 3σ limit |
| C III | 977 | <31 | … | … | <12.65 | [-400, -190] | 3σ limit |
| C II | 1036 | <11 | … | … | <13.0 | [-400, -190] | 3σ limit |
| N II | 1084 | <12 | … | … | <13.0 | [-400, -190] | 3σ limit |
| Si II | 1193 | <11 | … | … | <12.3 | [-400, -190] | 3σ limit |

[a] A limit is not reported for Si III λ1206 over the velocity range from -190 to -400 km s$^{-1}$ because of contamination from another absorber (see Fig. 2).

[b] The Voigt profile fit code of Fitzpatrick & Spitzer (1994) was used to obtain the O VI component fit results. The line spread function was taken from Ghavamian et al. (2009). The reduced $\chi^2$ of the O VI fit is 0.38.

[c] These 3σ limits for H I absorption at -278 km s$^{-1}$ are determined from the Voigt profiles displayed in Figure 9 for b = 60 km s$^{-1}$ (upper panel) and b = 150 km s$^{-1}$ (lower panel).

[d] The N V measurements listed here are discussed in detail in §5.1. Special efforts were required to deal with continuum placement uncertainties and the effects of COS fixed pattern noise for this weak absorber.

Table 5. Properties of the Plasma Associated with the Broad
H I-Free O VI Absorber for different values of $b_t$(O VI)

| $b_t$(O VI) (km s$^{-1}$) | $b_{nt}$(O VI) (km s$^{-1}$) | log T | log N(H$^+$) | log N(H I) | b(H I) | $\tau_0$(H I) |
|---|---|---|---|---|---|---|
| 25.6 | 45.3 | 5.80 | 19.11 | 12.75 | 102 | 0.042 |
| 32.3 | 40.8 | 6.00 | 19.71 | 13.08 | 129 | 0.071 |
| 36.8 | 36.8 | 6.11 | 19.87 | 13.10 | 146 | 0.065 |
| 52.0 | 0 | 6.41 | 20.50 | 13.41 | 207 | 0.094 |

[a] The estimates of log N(H$^+$) and log N(H I) assume a solar abundance for oxygen, log N(O VI) = 13.90, and CIE. Note that T = 961 $b_t$(O VI)$^2$, with $b_t$ in km s$^{-1}$. With the observed Doppler parameter for O VI, $b_o$ = 52 km s$^{-1}$, we have estimated $b_{nt}$ and $b_t$ by assuming the non-thermal broadening has a Gaussian distribution and using $b_{nt}^2 = b_o^2 - b_t^2$.

FIGURES

FIG. 1. COS (lower panel) and STIS (upper panel) spectra from 1195 to 1230 Å. The two O VI systems detected in H I $\lambda$1025 and O VI $\lambda\lambda$1032, 1037 at z = 0.16716 and 0.18295 are identified in the lower panel. The much higher S/N in the COS spectrum allows the unambiguous study of strong and weak O VI absorbers. Observed flux is plotted against heliocentric wavelength. PKS 0405-123 increased in brightness between the STIS and COS observations.

FIG.2 WFPC2 F702W image of the region surrounding PKS 0405-123 showing the two galaxies with redshifts of 0.1668 and 0.1670 having impact parameters of 116$h_{70}^{-1}$ and 99$h_{70}^{-1}$ kpc, respectively. The redshifts of other galaxies are also listed. The figure is adapted from Chen & Mulchaey (2009).

FIG. 3a and b. COS observations of continuum normalized flux versus restframe heliocentric velocity for various ions in the z = 0.16716 absorber over the velocity range from -450 to 225 km/s. The species are identified in each panel. Various contaminating IGM or ISM absorption lines are marked with x's.

FIG. 4. (upper panel) Si II $N_a(v)$ profiles are compared for Si II $\lambda\lambda 1190, 1193, 1260$ and 1304. There is decreasing line saturation in the sequence Si II 1260, 1193, 1190, and 1304. (lower panel) $N_a(v)$ profiles are compared for Si II $\lambda 1190$, N II $\lambda 1084$, and C II $\lambda 1335$.

FIG. 5. O VI $\lambda\lambda 1032$ and 1037 $N_a(v)$ comparison. The $N_a(v)$ profiles for the strong O VI $\lambda 1032$ and weak $\lambda 1038$ O VI lines are in excellent agreement over the velocity range from 100 to -325 km s$^{-1}$ which indicates there is no hidden unresolved saturation in the measurements. The increase in $N_a(v)$ for the O VI $\lambda 1038$ line for v < -325 km s$^{-1}$ is caused by the blending from the C II $\lambda 1036$ line in the z = 0.16716 system (see Fig.2).

FIG. 6. $N_a(v)$ comparison profiles for O VI $\lambda 1032$ with H I $\lambda 1216$, N V $\lambda 1243$, C III $\lambda 977$, Si III $\lambda 1207$, and Si II $\lambda 1190$. The $N_a(v)$ profiles for C III and Si III are strongly affected by unresolved saturation near the maximum of absorption. Although the high ion O VI and N V profiles are much broader than those for the moderately ionized gas,

they approximately track the velocity range of absorption seen in the moderately ionized gas.

FIG. 7 A simple single slab photoionization model for the moderately ionized gas in the z =0.16716 absorber. The radiation background is from the Haardt & Madau (2001) model interpolated to z = 0.16716 and includes ionizing photons from AGNs and star forming galaxies. log N(X) is plotted against the logarithm of the photoionization parameter log U. Heavy solid lines on the lighter curves for the various ions display the observed column density range. For many species that range is large because the observations only provide a lower or upper limit. The elemental abundances adopted in the model are [O/H] = 0.1, [N/H] = 0.40, [C/H] = 0.0, [Si /H] = -0.4 and [Fe/H] = 0.0. The complexity of the absorber and the assumption of a simple single slab model makes it difficult to obtain accurate information from the modeling.

FIG. 8. (left panel) Voigt profile fits to the broad H I-free O VI $\lambda$1032 absorption measured by COS assuming a single broad component. The component parameters derived are v = -278±3 km s$^{-1}$, b = 51±1km s$^{-1}$, and log N(O VI) = 13.90±0.03. The same parameters do a fair job of explaining the absorption behavior over the velocity range from -200 to -300 km s$^{-1}$ for the O VI $\lambda$1038 absorption. For v < -300 km s$^{-1}$ the O VI $\lambda$1037 absorption is contaminated by C II $\lambda$1036 at z = 0.16716 (see Figs. 3 and 5). (right panel) The broad H I-free O VI $\lambda\lambda$1032,1038 absorption recorded in the lower S/N STIS observations binned to 10 km s$^{-1}$. The profile displayed on the STIS spectrum is that obtained from the fit to the COS observations.

FIG. 9. The normalized flux from 0.5 to 1.0 to each side of the $z = 0.16716$ Ly $\alpha$ absorption is displayed over a velocity range of $\pm 1000$ km s$^{-1}$. The upper panel displays Voigt absorption profiles of H I absorption at -278 km s$^{-1}$ with $b = 60$ km s$^{-1}$ and log N(H I) = 12.7, 13.0 and 13.3. The lower panel shows H I absorption with $b = 150$ km s$^{-1}$ and log N(H I) = 13.1, 13.4, 13.7. The Voigt profiles have been convolved with the COS LSF. The 3$\sigma$ upper limits to H I absorption associated with the broad O VI absorber are estimated to be log N(H I) < 12.7 for $b = 60$ km s$^{-1}$ and log N(H I) < 13.1 for $b = 150$ km s$^{-1}$.

FIG. 10. Photoionization model for the broad H I-free O VI absorber. The radiation background is from the Haardt & Madau (2001) model interpolated to $z = 0.16716$ and includes ionizing photons from AGNs and star forming galaxies. The hydrogen column density [log N(H I)= 12.7] in the absorber was chosen to be the observed upper limit. An oxygen abundance of [O/H] ~ 0 is required to explain the large observed O VI column density of log N(O VI) = 13.90. If the true amount of H I in the absorber is smaller than the 3$\sigma$ upper limit, a larger oxygen abundance would be required to produce the observed amount of O VI. The 'fit' shown for [O/H] = 0 requires log U = -0.2, a path length of 180 kpc through gas with log n(H) = -5.5, log T = 4.72 and P/k = 0.4 cm$^{-3}$ K. No measurements exist for C IV and Ne VIII. The limit log N(Si IV) < 12.43 is satisfied by the model. The column density log N(N V) = 13.16±0.12 would require a 0.3 dex enhancement in [N/O] to be consistent with the model.

FIG. 11. Column densities versus log T for highly ionized ions and H I assuming solar abundance ratios and collisional ionization equilibrium with log N(H) = 19. For log T from 5.5 to 6.5 the amount of O VI is 2 to 2.5 dex less than for O VII. However, COS can easily detect O VI column densities of log N(O VI) > 13.5 in broad absorption lines with b ~ 35 to 60 km s$^{-1}$ at a resolution of ~17 km s$^{-1}$. With current X-ray satellites it is difficult to detect O VII columns smaller than log N(O VII) ~ 16.0 at a resolution of ~700 km s$^{-1}$.

FIG. 12. Log [N(O VI)/N(H I)] vs log U for a hybrid model of photoionization and collisional ionization in gas with T = $10^5$, $3.2 \times 10^5$ and $10^6$ K and solar abundances for O/H. The photoionizing background is that for z = 0.167 from Haardt & Madau (2001) and includes ionizing photons from AGNs and star forming galaxies. For small log U (left side of the figure) the value of log [N(O VI)/N(H I)] approaches the value found in CIE at the given temperature. The changes for large U (right side of the figure) show the increasing effects of the ionizing background. At T = $10^6$ K the ionizing background is not important until log U > -1 corresponding to $n_H$ < $10^{-5}$ cm$^{-3}$ where the modification is -0.13 dex. The ion ratio model curves for a solar oxygen abundance agree with the observed limits for log [N(O IV)/N(H I)] where the curves are shown with the heavy lines.

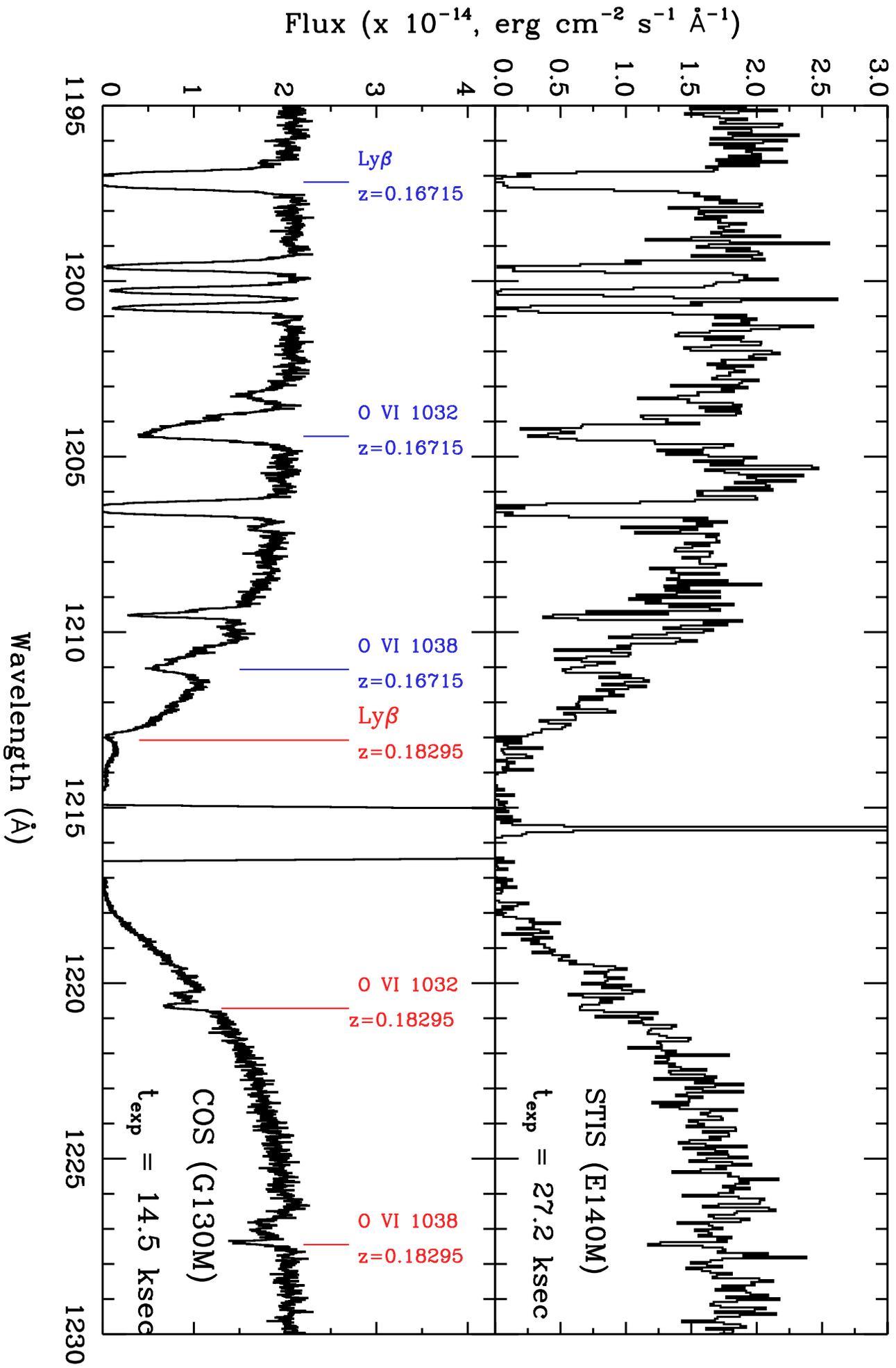

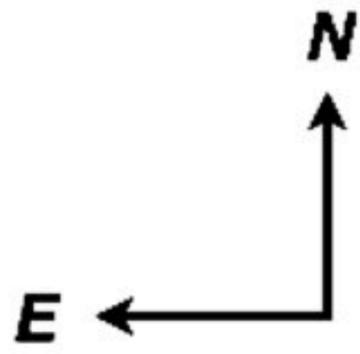

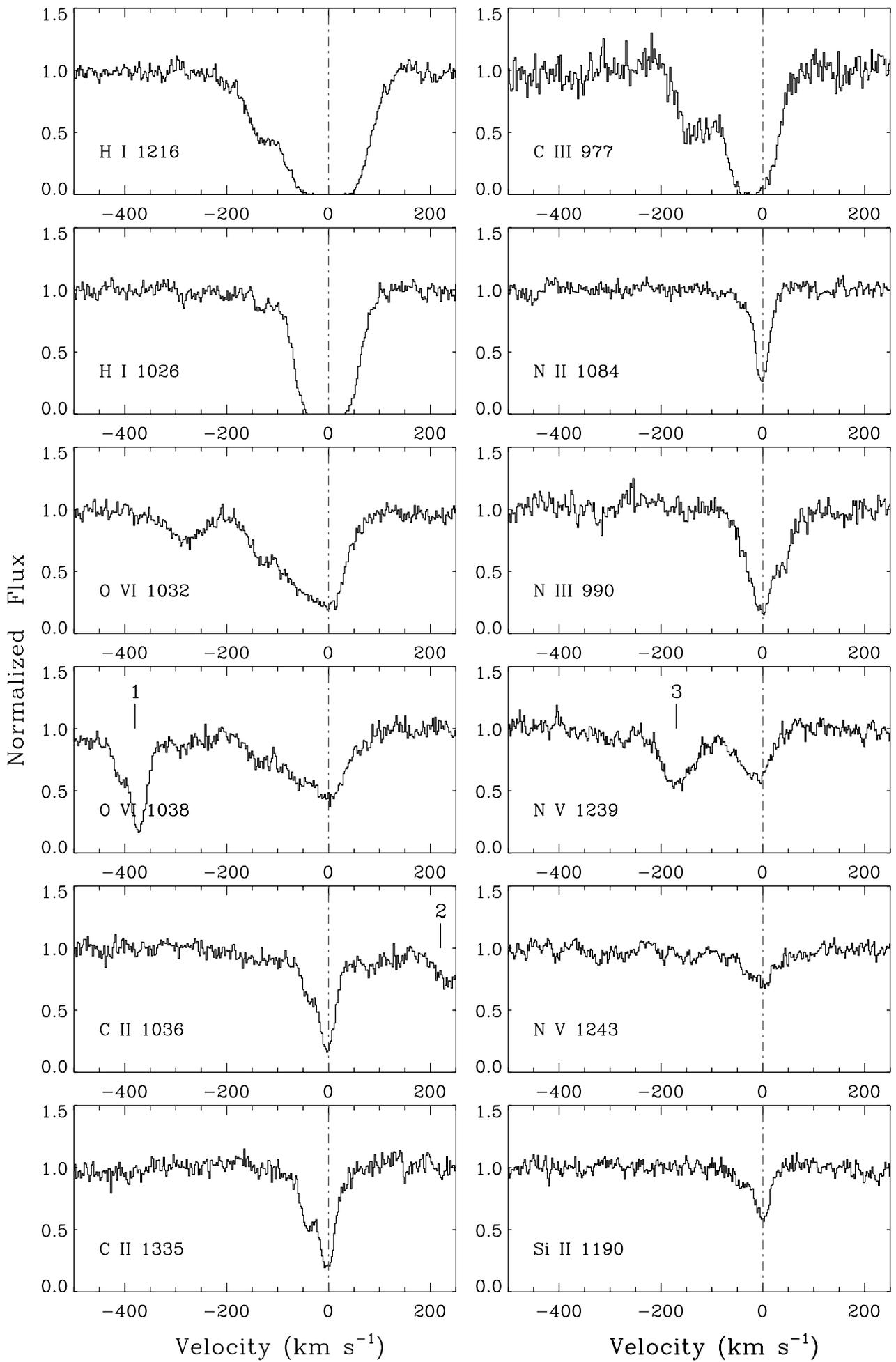

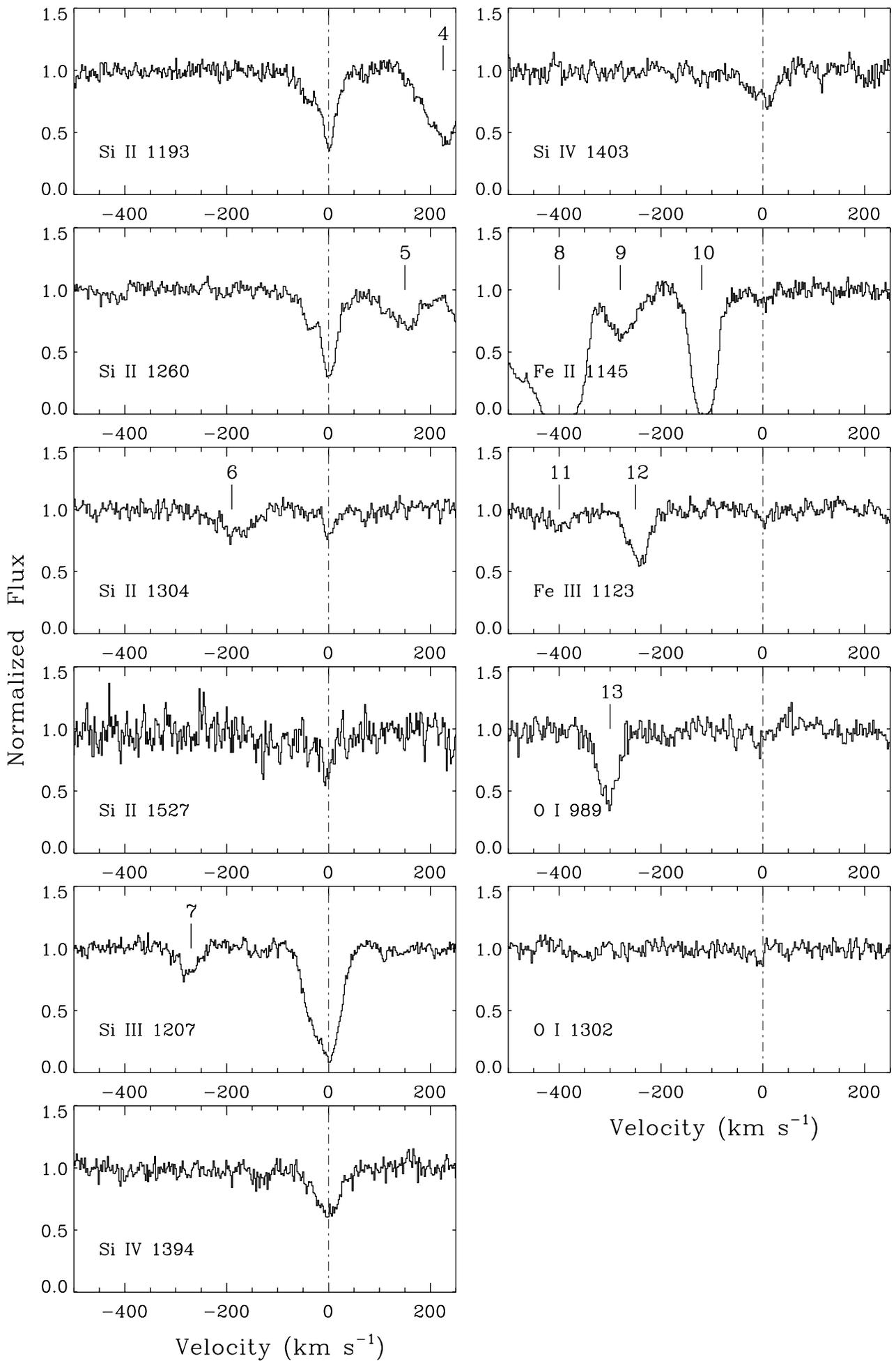

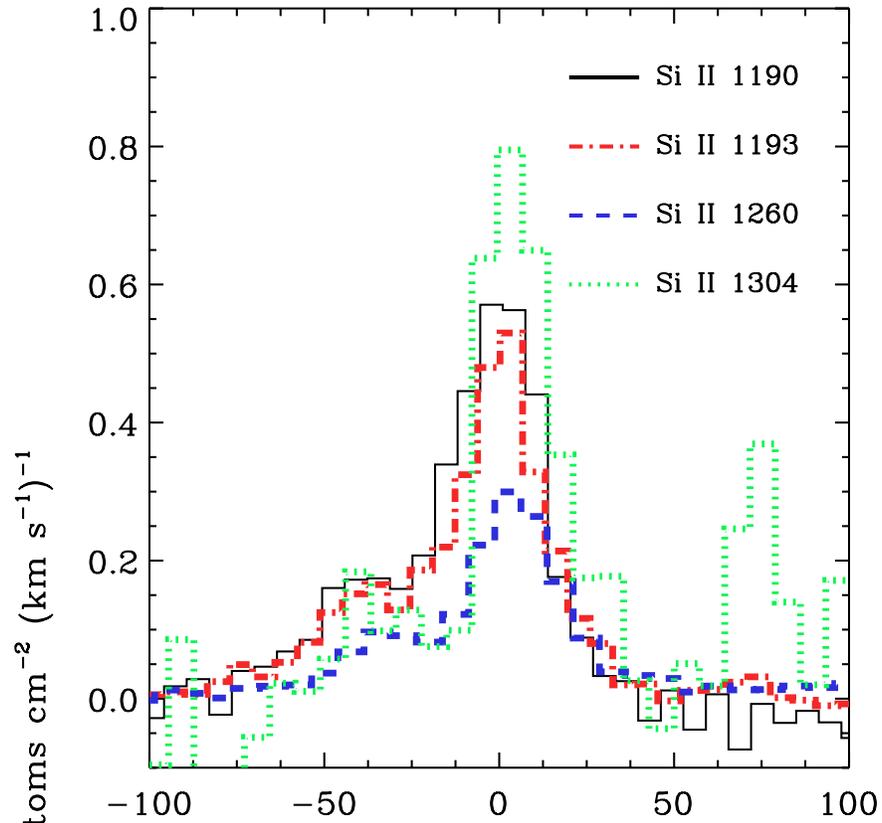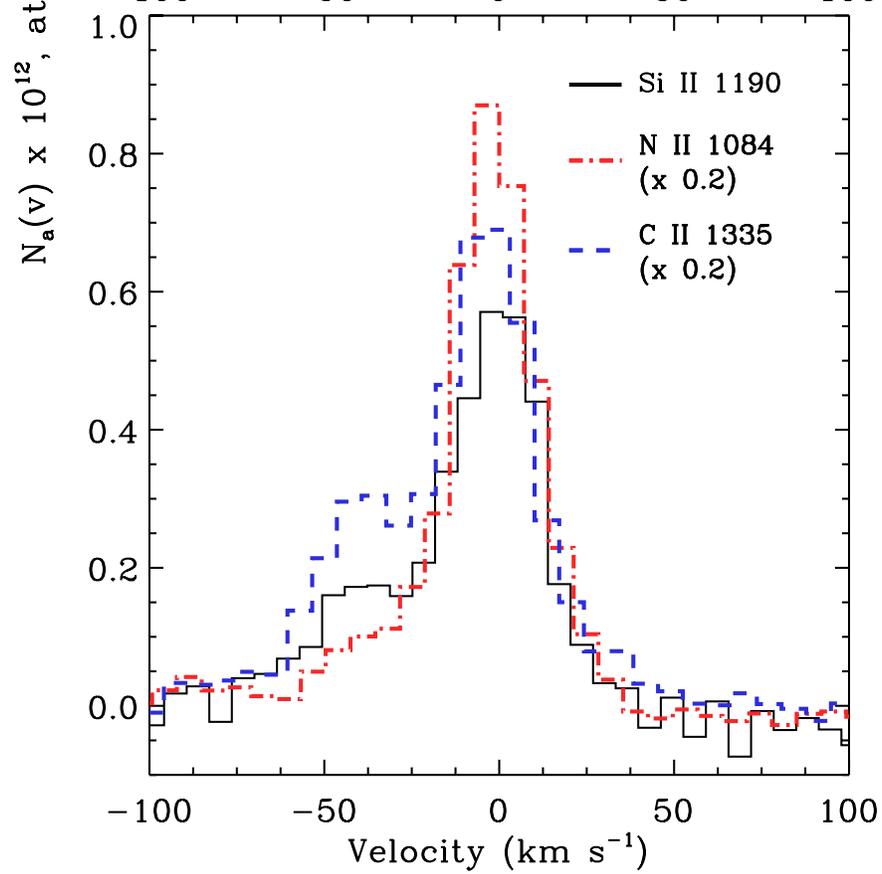

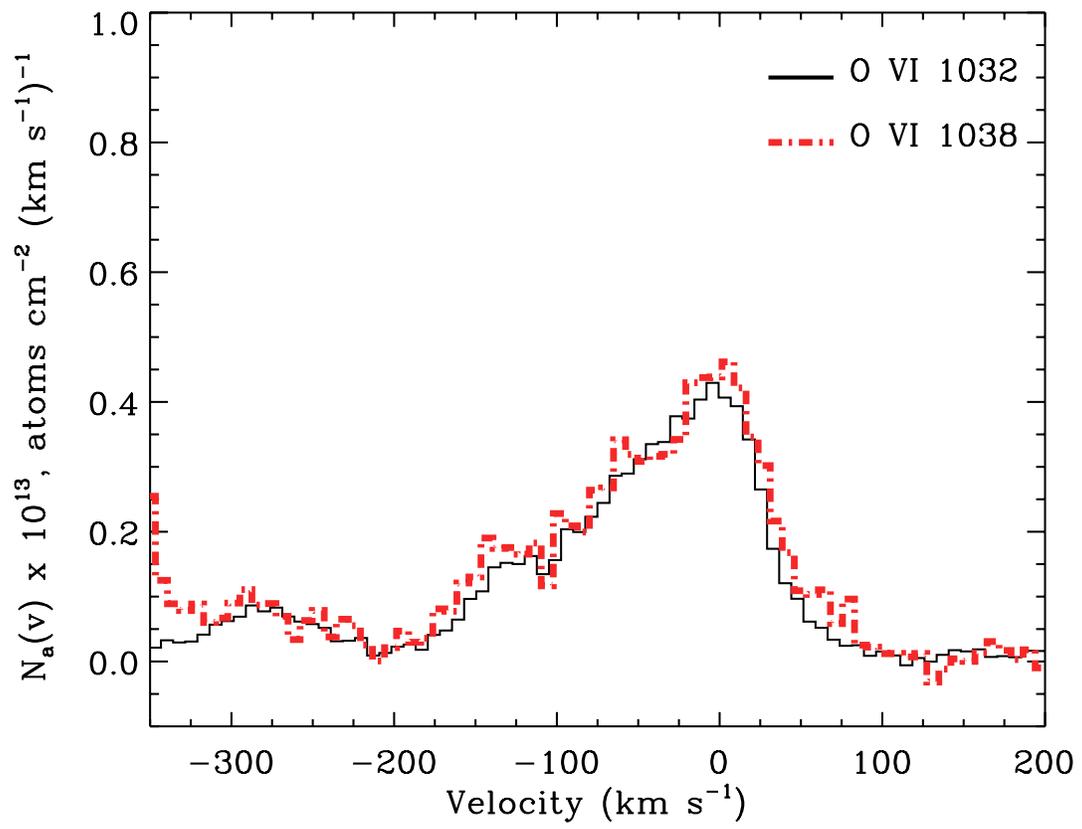

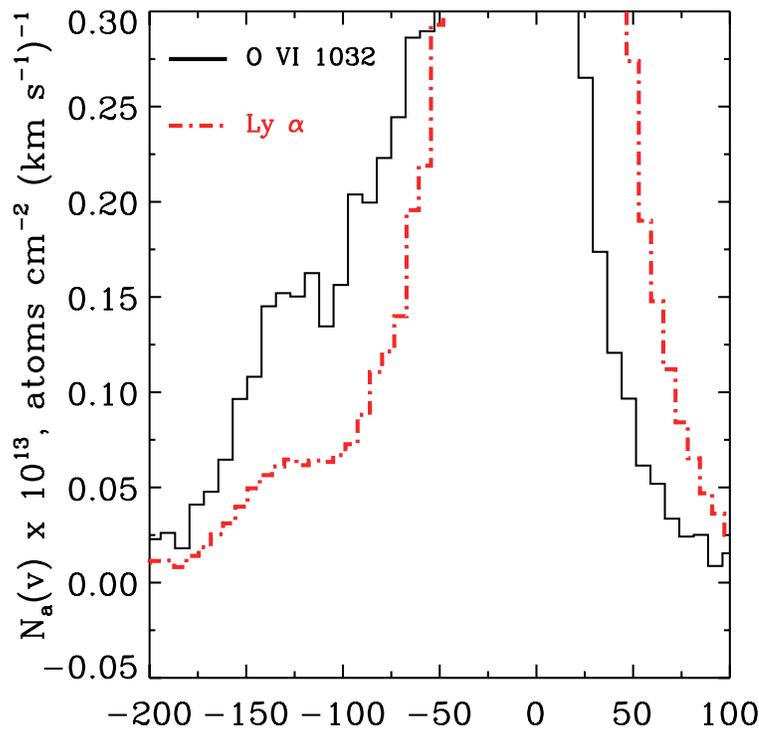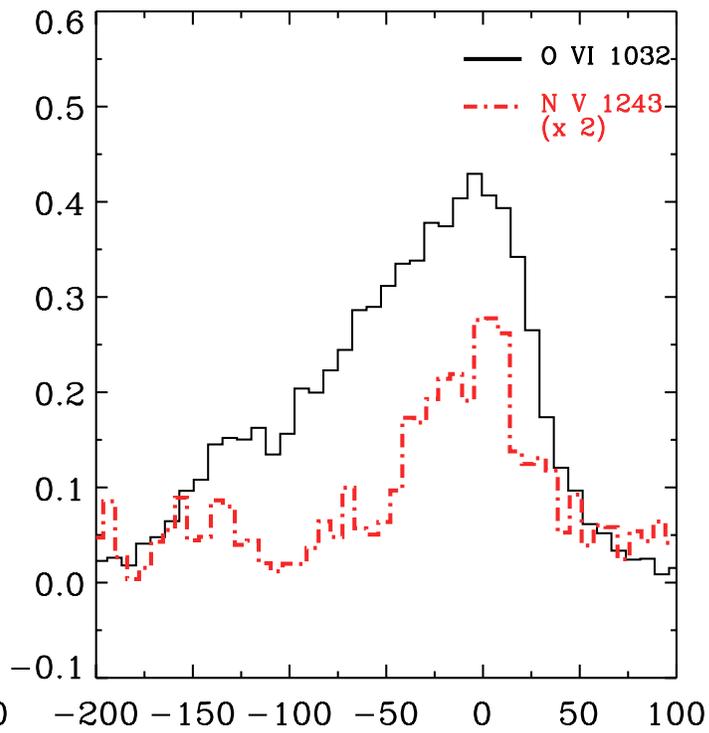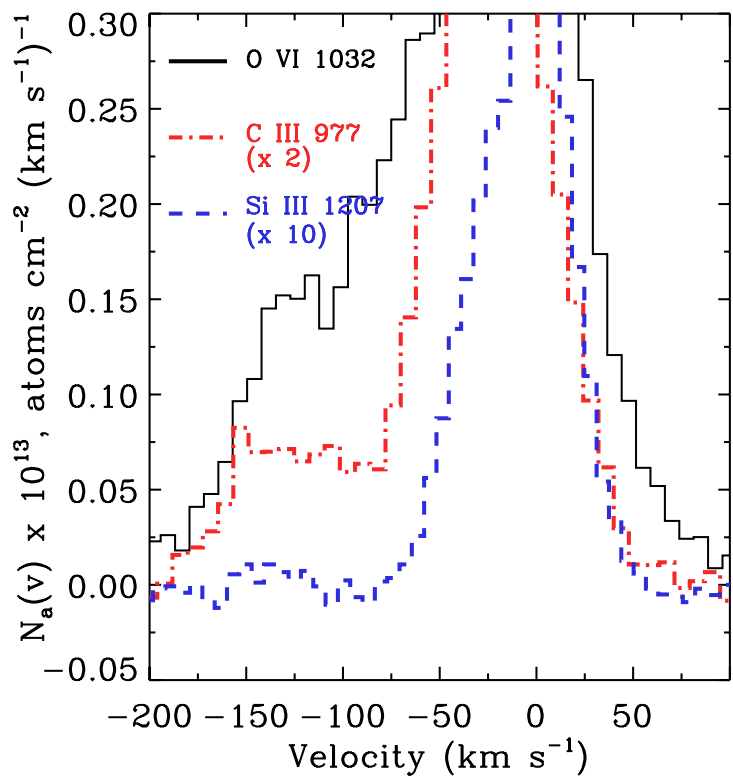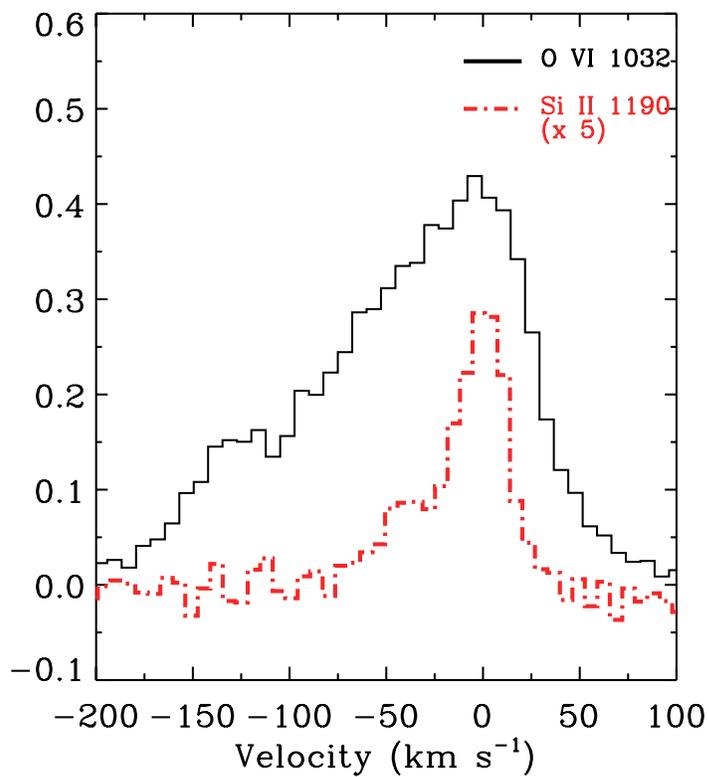

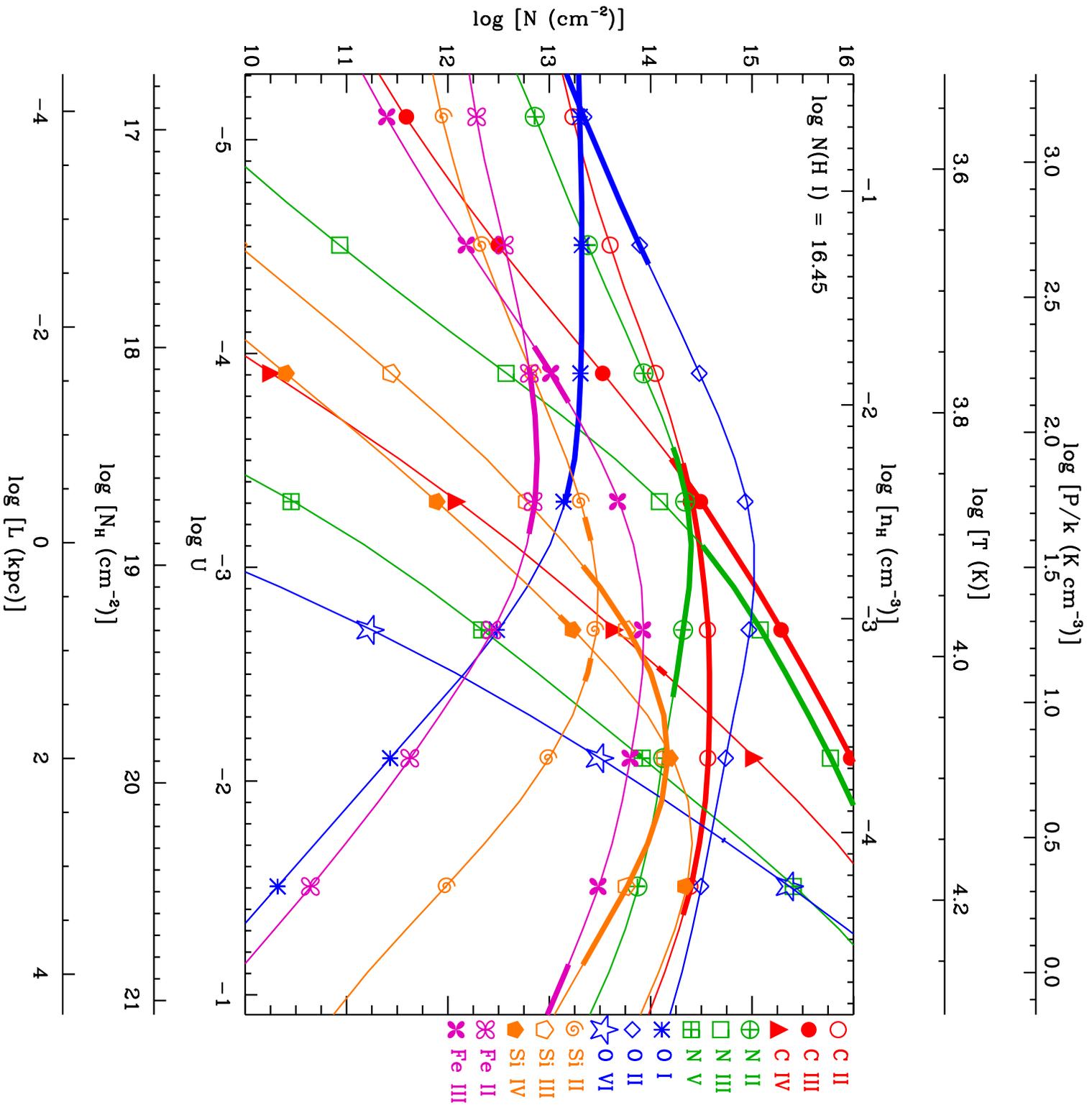

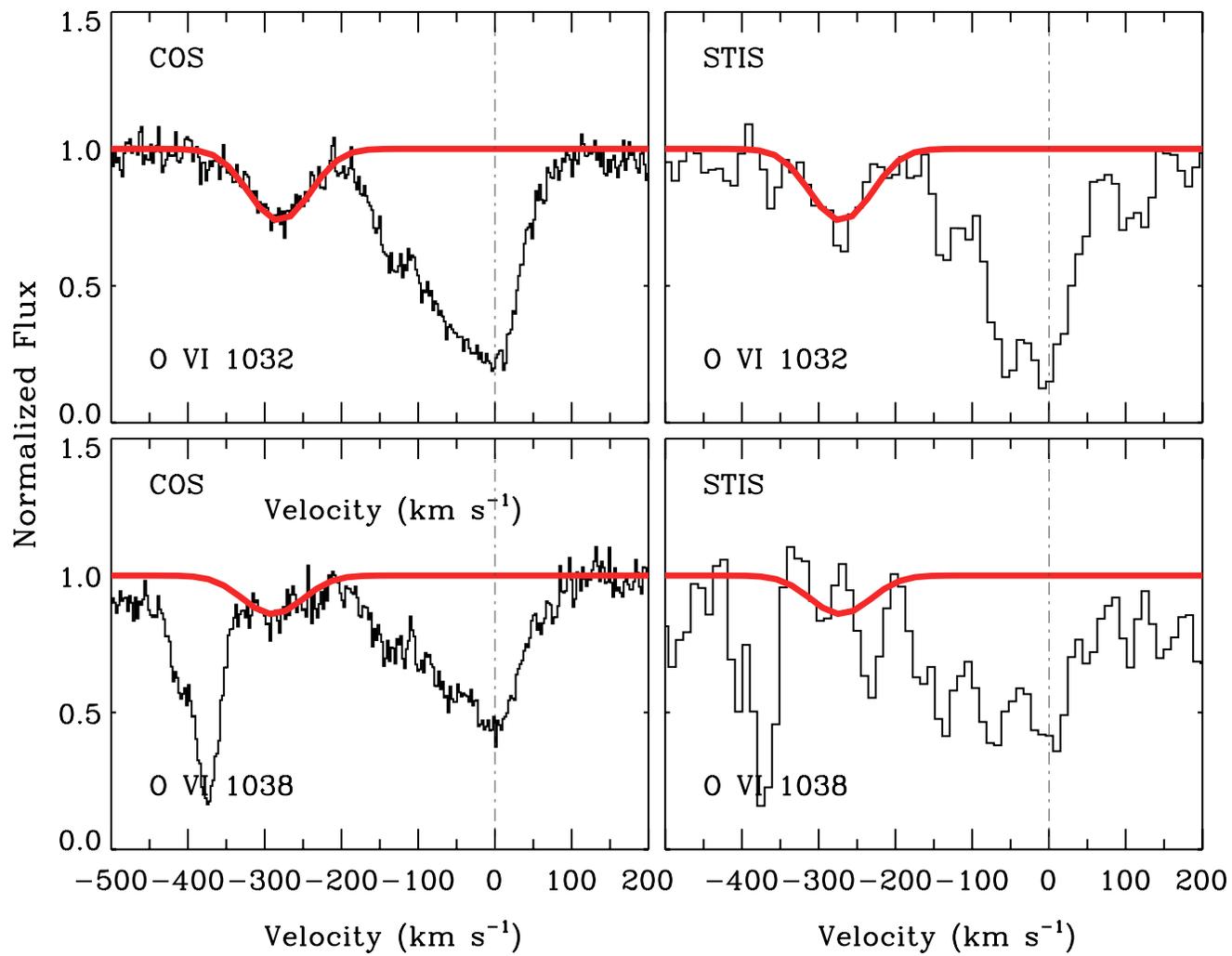

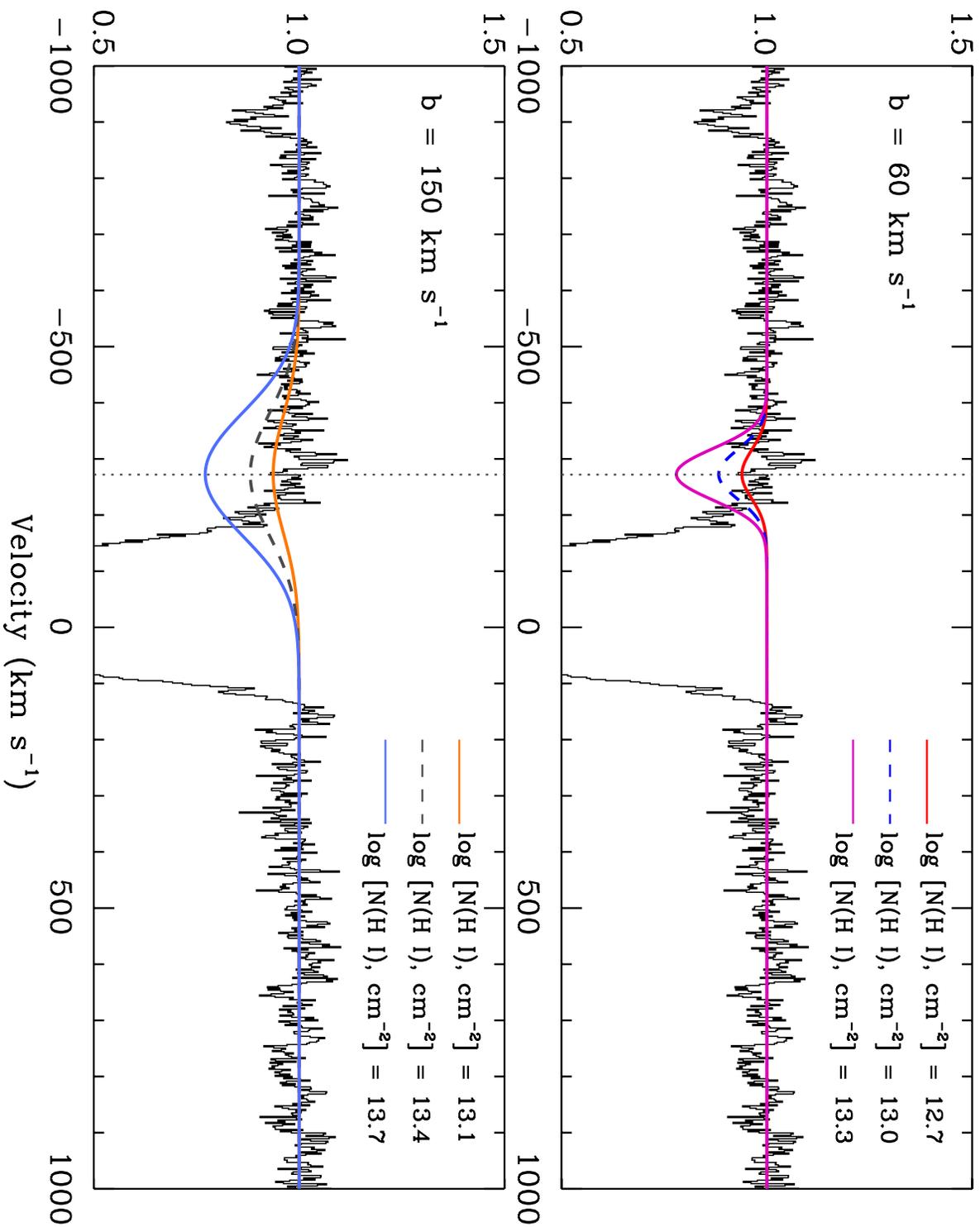

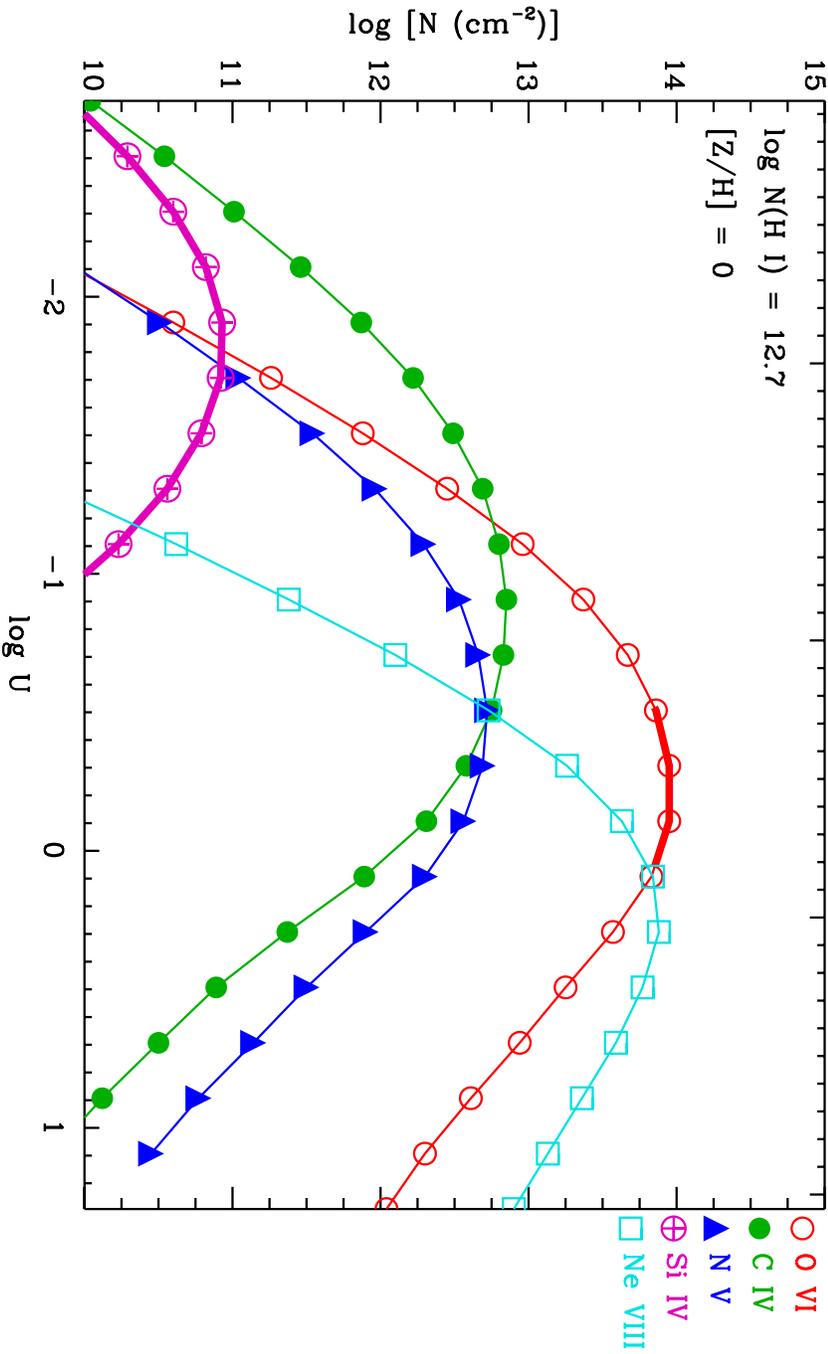

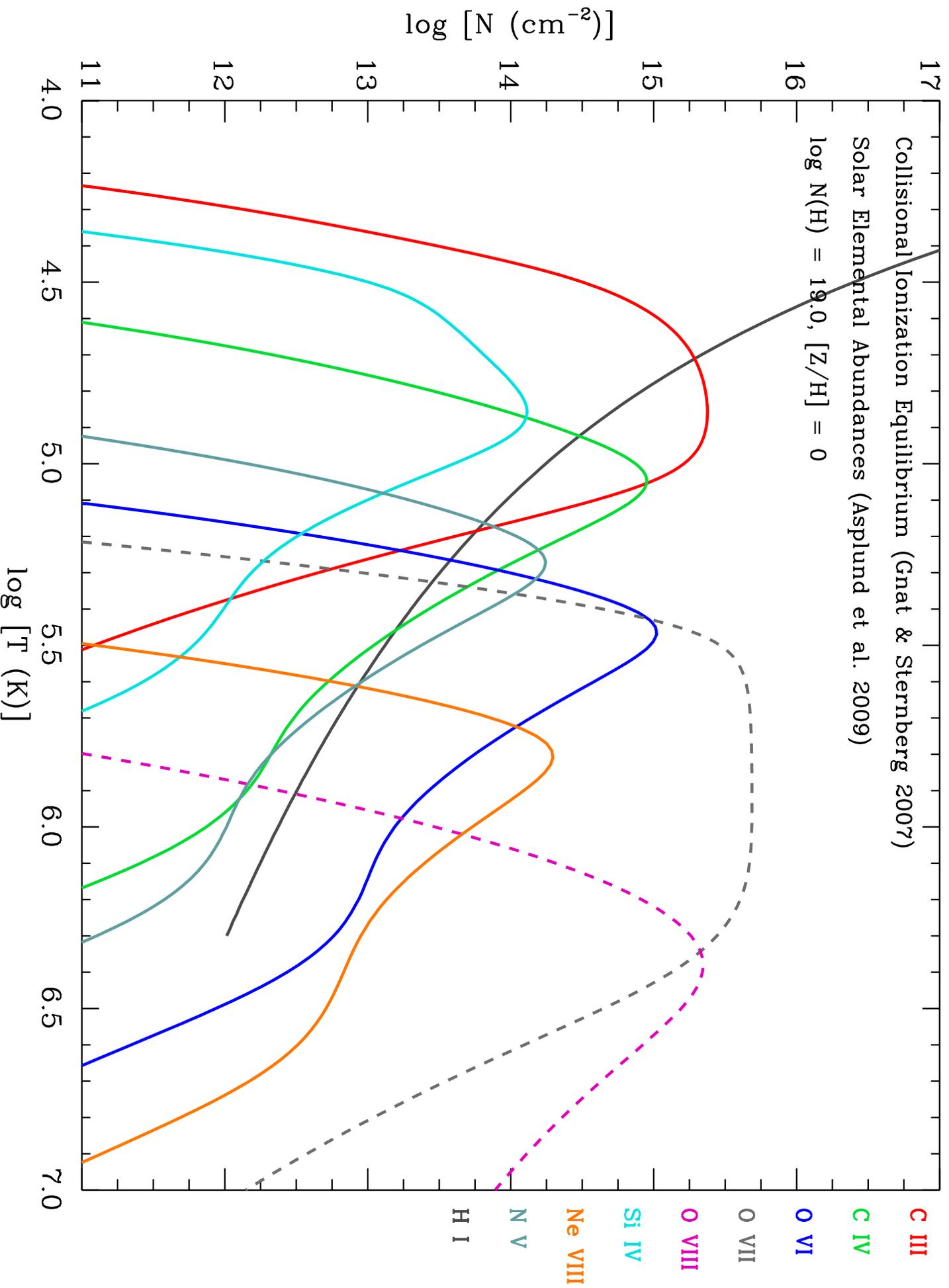

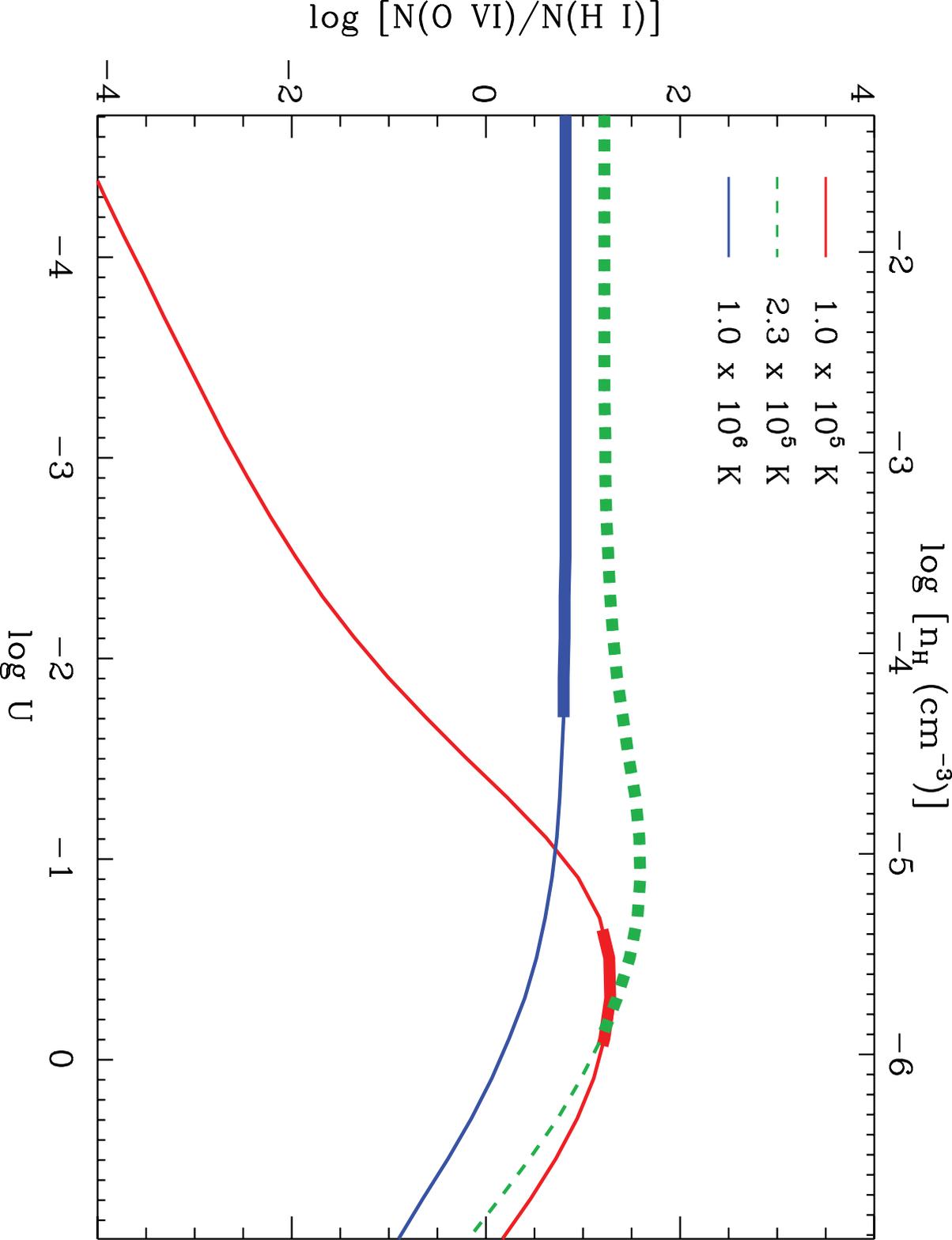